\documentclass{jfm}
\usepackage{lmodern}
\usepackage{bm,amsfonts,amssymb,amsmath}
\usepackage{graphicx,color,array,tabularx}
\usepackage{newtxtext,newtxmath}
\usepackage{epstopdf,epsfig}
\usepackage{natbib}
\usepackage{hyperref}
\usepackage[width=\textwidth,justification=justified]{caption}

\usepackage{tikz}
\usetikzlibrary{arrows,arrows.meta,calc}
\usepackage{pgfplots}
\pgfplotsset{height=8cm,compat=1.16}

\definecolor{rouge}{RGB}{178, 33, 51}

\title{Sheared stratified turbulence driven by Kolmogorov flow}

\author{
Alessandro Sozza\aff{1,2}\corresp{\email{alessandro.sozza@cnr.it}} and 
Andrea Maffioli\aff{3}
}

\affiliation{
\aff{1} Institute of Atmospheric Sciences and Climate, National Research Council, Corso Fiume 4, 10133, Torino, Italy.
\aff{2} Laboratoire de Physique, \'Ecole Normale Sup\'erieure de Lyon, Universit\'e de Lyon, CNRS, 46 All\'ee d'Italie, 69342, Lyon, France.
\aff{3} \'Ecole Centrale de Lyon, CNRS, Universit\'e Claude Bernard Lyon 1, INSA Lyon, LMFA, UMR5509, 69130, \'Ecully, France
}

\begin{document}

\maketitle

\begin{abstract}
We investigate three-dimensional turbulence in a stably stratified fluid driven by a vertically sheared Kolmogorov flow using direct numerical simulations of the Boussinesq equations. As stratification increases, mean profiles evolve toward piecewise-linear shapes while layered density structures emerge, with sharp interfaces separating well-mixed bulk layers. These highly stable interfaces form in the low-shear regions of the mean velocity profile and tend to promote flow relaminarisation, while shear-generated turbulence persists in the bulk layers. We analyse turbulent fluctuations, buoyancy transport and its spatial organisation, and flow stability via profiles of the gradient Richardson number $Ri_g$. The Richardson number in the bulk layers remains of order unity or less, $Ri_g \lesssim 1$, so that efficient turbulent shear production can take place there. Mixing efficiency analysis shows that the Nusselt number scales with the buoyancy Reynolds number $Re_b$ as $Nu = 1 + \Gamma Re_b$ (with $\Gamma = \epsilon_p / \epsilon$), with the data collapsing onto a robust master curve and roughly following a power-law $Nu \sim Re_b^{0.8}$. Further increase of stratification leads to a temporally intermittent turbulent regime, characterised by quasi-periodic bursts. We propose that the transition from stationary turbulence to this temporally intermittent regime is controlled by the buoyancy Reynolds number and highlight the mechanisms disrupting the turbulence and layered structures.
\end{abstract}



\section{Introduction}
\label{sec:intro}

Geophysical and astrophysical flows occur at large scales and their consequently high Reynolds number means they are almost invariably turbulent. A stable density stratification is additionally a common feature of such flows, which shapes their structure and alters their dynamics. The vast majority of the Earth's ocean is stably stratified and this feature often has an order-one influence on the dynamics of the ocean \citep[][]{thorpe2007}. In the ocean, at relatively small scales (horizontal scales $\lesssim 1$ km), the stratification is the dominant body force and the Coriolis force due to the Earth's rotation is subdominant \citep[][]{kunze2019}. This means that to study this range of oceanic scales, and a corresponding range of scales in the atmosphere, we can focus on non-rotating stratified turbulence. 

In this study we focus our attention on a linearly stratified fluid with a constant Brunt-V\"ais\"al\"a frequency $N$, forced by a horizontal body force, which generates a mean flow with a vertical shear. Our general objective is to study the interaction between the mean vertical shear, the density stratification and the turbulence. The present configuration in which turbulence is fed by a mean vertical shear, in a vertically stratified fluid, has received considerable attention in the literature because it constitutes a simple "test case" that replicates some phenomena --- vertical shear instability and breakdown to 3D turbulent motions --- that are ubiquitous in geophysical fluid dynamics. We herein refer to this vertically sheared configuration as a \textit{stratified shear flow}; this expression in general encompasses a larger class of flows, including horizontally sheared flows \citep[see, e.g.,][]{lucas2017}, but we here restrict our attention to flows with purely vertical shear.

There is an important existing literature on stratified shear flows. Concentrating on the recent work making use of direct numerical simulations (DNS), \citet{salehipour2015,salehipour2016,smith2021} have considered the dynamics of isolated mixing layers. In their high Reynolds number DNS, they simulated the canonical configuration of co-located shear and density gradient. This configuration is close to a horizontally flowing two-layer fluid, in which the vertical shear of the horizontal flow is concentrated at the density interface. At high Reynolds number, the density interface undergoes shear instability and breaks down to turbulence. This setup is relevant for two-layer exchange flows, as found, for example, at river confluences \citep{dureuil2025} or in estuarine flows \citep{farmer2002}. The numerical studies of \citet{salehipour2015,salehipour2016,smith2021} focused on the turbulent mixing of the density field, on the associated mixing efficiency and on how it varied depending on the important non-dimensional parameters and on the type of shear instability, Kelvin-Helmholtz instability \citep[][]{salehipour2015} or Holmboe instability \citep[][]{salehipour2016}. A somewhat similar configuration has been considered in the experiments of \cite{lefauve2019,duran2023}, in which a two-layer exchange flow was created in a long rectangular duct, which had its ends immersed in two basins filled with salt-water solutions of different density. An exchange flow between the two basins was created in this apparatus by tilting the duct at a small angle with respect to the horizontal. Depending on the tilt angle, different flow regimes were observed: laminar flow, mostly laminar flow with travelling Holmboe waves, spatio-temporally intermittent turbulence with alternating laminar and turbulent phases, and finally sustained turbulence \citep{lefauve2019}. \citet{lefauve2019} found that for high enough tilt angles, the regime transitions were controlled by a parameter given by the Reynolds number times the tilt angle. This parameter should asymptotically become proportional to the buoyancy Reynolds number $Re_b$ \citep{lefauve2019}.

Another important class of stratified shear flows is represented by wall-bounded stratified turbulent flows. Previous numerical studies have investigated stratified channel flow \citep[][]{garg2000,zonta2012,cen2024}, stratified Couette flow \citep[][]{zhou2017}, as well as stratified open-channel flow \citep[][]{issaev2023}. These studies have explored a wide range of phenomena, including turbulent mixing \citep[][]{zhou2017}, the onset of intermittent turbulence \citep[][]{issaev2023,cen2024}, and the effect temperature-dependent fluid properties \citep[][]{zonta2012}. In such wall-bounded configurations, stable stratification is not directly imposed but rather arises as a consequence of the thermal boundary conditions at the walls. These consist either of fixed temperature boundary conditions or of fixed heat flux boundary conditions at the walls. 

Several of the aforementioned studies have reported the formation of relatively well-mixed regions in the flow, separated by sharp, stably stratified density interfaces. For instance, in the stratified channel flow simulations by \cite{zonta2012}, a stable interface develops at the channel mid-height, separating two well-mixed regions. Similar findings were reported by \citet{cen2024}, who observed that the density interface at mid-height strengthens with increasing stratification. The channel mid-plane corresponds to a location of zero mean vertical shear, due to the flow symmetry, so that the density interface in stratified channel flow emerges precisely in a region of vanishing vertical shear. In contrast, in stratified Couette flow, no distinct density interface forms, since this configuration is more constrained than stratified channel flow, as the mean vertical shear is non-zero throughout the entire height of the domain \citep[see the discussion in][]{zhou2017}. 
Isolated mixing layers and exchange flows \citep[e.g.][]{salehipour2015,lefauve2019} feature a density interface that, by construction, coincides with the region of maximum vertical shear. This motivates the exploration of more idealized configurations of stratified shear flow, in which the evolution of the mean density profile is subject to fewer geometric or dynamical constraints. For such flows, a key issue concerns the relative positioning of the regions of strongest stratification and strongest shear. When these locations are not prescribed, the density interface—identified as the region of strongest density gradient—may emerge at a position that is dynamically determined by the non-linear interactions within the flow. Its location and structure therefore result from the self-organization of the stratified turbulence. In this work we are interested in studying this self-organization for a chosen configuration of stratified shear flow.

A convenient framework for investigating such idealised shear-driven dynamics is provided by the Kolmogorov flow, in which a steady horizontal body force varies sinusoidally in the vertical direction. This setup has been extensively employed in DNS of sheared turbulence \citep[][]{borue1996,musacchio2014,lalescu2021}. Over the years, it has also served as a versatile model for exploring additional physical processes, such as the interaction between turbulence and suspended particles in dusty or multiphase flows \citep[e.g.][]{delillo2016,sozza2020,sozza2022}, the dynamics of active or gyrotactic swimmers under shear \citep[e.g.][]{santamaria2014,borgnino2022}, and the role of viscoelasticity in polymer-laden turbulence \citep[e.g.][]{boffetta2005,garg2021}. These studies collectively highlight the flexibility of the Kolmogorov setup as a minimal yet powerful model to probe the interplay between shear, forcing, and additional physical mechanisms. The absence of solid boundaries further allows one to disentangle the intrinsic flow dynamics from wall effects, while the periodic nature of the forcing makes the setup particularly well suited for implementation in triply periodic pseudo-spectral codes. 

In geophysical and astrophysical contexts, stratified Kolmogorov flows have attracted moderate attention. Seminal work by \citet{balmforth2002,balmforth2005} analysed the linear and weakly non-linear stability, identifying key instability mechanisms and parameter regimes near onset. DNS studies \citep{garaud2015,garaud2016} extended this work to high Reynolds numbers and strongly stratified regimes, focusing on low Prandtl number (i.e. $Pr \le 0.1$) relevant for astrophysical flows, for which thermal diffusion plays a dominant role. They reported rich dynamics, revealing transitions from sustained steady-state turbulence to intermittent turbulent bursts alternated with long quiescent intervals of flow relaminarization. Such quasi-periodic bursting events have also been observed in the stratified inclined duct experiments of \citet{lefauve2019} and may be a general feature of stratified shear flows when subject to strong stratification. Later works considered horizontally sheared configurations of Kolmogorov flows \citep[e.g.][]{lucas2017,cope2020} showing that despite geometric differences, similar dynamical regimes and phenomenological features -- such as layer formation and intermittent bursting -- can emerge.

In the present study, we investigate the turbulent stratified Kolmogorov flow as a minimal model to study the self-organization of stratified shear turbulence. Our primary objective is to address the literature gap at $Pr=1$ and to determine whether the mean flow structures and dynamical regimes observed by \citet{garaud2016} at lower Prandtl numbers persist in this configuration. A second objective is to characterize the statistically stationary turbulent regime, which has not been previously analysed in detail. This includes the mean vertical profiles of velocity and density and their dependence on the governing parameters, the emergence and vertical scale of density interfaces, the characteristics and scaling of turbulent fluctuations, and the vertical transport of buoyancy, including mixing efficiency and Nusselt number. Finally, we examine the transition to the intermittent bursting regime, identifying parameter regimes in which quasi-periodic bursts occur and relating our findings to analogous observations in previous studies.

\section{Model}
\label{sec:model}

\subsection{Governing equations and basic phenomenology}
\label{subsec:eqs}

We consider a three-dimensional stably stratified fluid characterized by a linear background density profile $\rho_0-\gamma z$ with mean gradient $\gamma$, such that the total density field is expressed by $\rho(\bm{x},t) = \rho_0 - \gamma z + \rho'(\bm{x},t)$, where $\rho'$ represents the density perturbation away from the background profile. For convenience, we rescale the density perturbation by the mean gradient as $\rho' = \gamma \theta$. Using this formalism, $\theta$ has the dimension of a length and can be interpreted as the isopycnal displacement field, where the displacement is calculated with respect to the isopycnal position in the unperturbed state, $\rho = \rho_0-\gamma z$. For simplicity, though, in the remainder of the paper we will refer to both $\theta$ and $\rho'$, interchangeably, as the density (or scalar) perturbation.

\begin{figure}
\centering
\resizebox{0.6\textwidth}{!}{
\begin{tikzpicture}
\draw[thick] (0,0) -- (8,0) -- (8,8) -- (0,8) -- cycle;
\draw[thick] (0,8) -- (5,11) -- (13,11) -- (8,8);
\draw[thick] (8,0) -- (13,3) -- (13,11);
\draw[dashed] (0,0) -- (5,3) -- (13,3);
\draw[dashed] (5,3) -- (5,11);
\draw[thick,->,red!70!orange,ultra thick] (0,0) -- (2,0) node[below] {\Large $x$};
\draw[thick,->,red!70!orange,ultra thick] (0,0) -- (0,2) node[left] {\Large $z$};
\draw[thick,->,red!70!orange,ultra thick] (0,0) -- (1.25,0.75) node[above left] {\Large $y$};
\node[blue] at (6,1.8) {\Large $\bm{f}(z)$};
\begin{axis}[axis line style={draw=none}, tick style={draw=none}, no markers, xtick=\empty, ytick=\empty, ymin=0,ymax=1, samples=1000, y=8cm]
\addplot[blue,ultra thick] ({2*cos(deg(2*pi*x))},x);
\addplot[blue,dashed,ultra thick](0,x);
\foreach \x in {0,0.07,...,1} \addplot[thick,blue,->,shorten >=1pt] coordinates {(0,\x) (2*cos(deg(2*pi*\x)),\x)};
\end{axis}
\draw[red!70!orange,ultra thick,dashed] (9.875,1.125) -- +(0,8) node (M1) {} ;
\draw[red!70!orange,ultra thick] (12.375,2.625) -- (10.5,9.5);
\foreach \x in {1,...,7} \draw[red!70!orange,thick] (M1.center) ++ (0,-\x) -- +(0.625*3*\x/8+0.625,0.375*3*\x/8+0.375);
\node[red!70!orange,rotate=31] at (10.7,5.15) {\Large $\rho_0 - \gamma z$}; 
\end{tikzpicture}
}
\caption{Schematic of system under study. The numerical domain of the main DNS runs is a cube of size $L=2\pi$, on which the external force $\bm{f}(z)$ and the linear background density profile $\rho_0-\gamma z$ are shown. The external force is steady in time and will generate a mean flow $\overline{\bm{u}}(z)$ oriented in the same direction (the $x$-direction) and with a similar modulation along the $z$-direction.}
\label{fig:schema}
\end{figure}
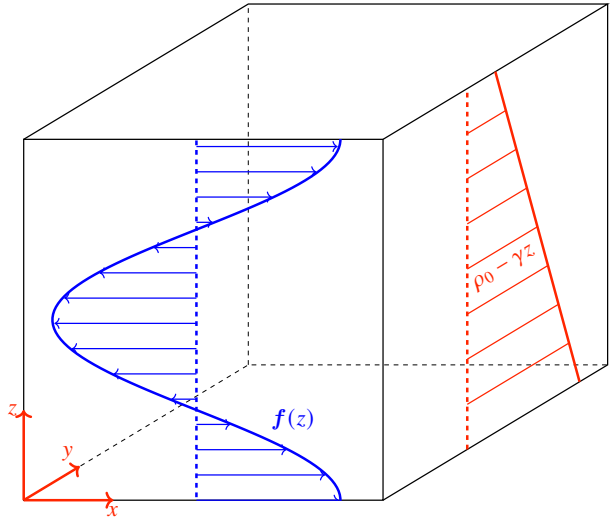

Within the Boussinesq approximation, the equations of motion of the velocity field $\bm{u}(\bm{x},t)$, with components $(u,v,w)$, and the density perturbation $\theta(\bm{x},t)$ are given by 
\begin{eqnarray}
&& \partial_t \bm{u} + \bm{u} \cdot \bm{\nabla} \bm{u} = 
- \bm{\nabla} p - N^2 \theta \widehat{\bm{z}} + \nu \nabla^2 \bm{u} + \bm{f}, \label{eq1} \\[0.2cm]
&& \partial_t \theta + \bm{u}\cdot\bm{\nabla} \theta = w +\kappa \nabla^2 \theta \label{eq2},
\end{eqnarray}
where $\nu$ is the kinematic viscosity, $\kappa$ is the diffusivity, $N^2=\gamma g/\rho_0$ is the squared Brunt-V\"ais\"al\"a frequency. The system is driven by an external forcing, known as Kolmogorov flow, which is represented by a monochromatic sinusoidal shear flow $\bm{f} = F\cos(Kz) \widehat{\bm{x}}$. A schematic of the present configuration is given in Figure~\ref{fig:schema}.

In the laminar regime, Eqs.~\eqref{eq1}--\eqref{eq2} admit a simple stationary solution given by $\bm{u} = U_0 \cos(Kz) \widehat{\bm{x}}$ with $U_0=F/(\nu K^2)$. In the absence of stratification, for $N=0$, this solution is known to become unstable to transverse large-scale perturbations when the Reynolds number $Re=U_0/(\nu K)$ exceeds the critical threshold $Re_c > \sqrt{2}$ via a long-wave instability mechanism \citep{meshalkin1961,sivashinsky1985}. Non-linear evolution of the instability leads to a cascade of modes and transition to turbulence. Surprisingly, the mean flow maintains a monochromatic velocity profile similar to the laminar one, although with a reduced amplitude. The persistence of this large-scale pattern in the turbulent state has been confirmed by several studies \citep{borue1996,musacchio2014,delillo2016,sozza2020}.

When stable stratification is present ($N > 0$), the behaviour of the flow is strongly modified. A classical result by \citet{miles1961} and \citet{howard1961}, known as the Miles–Howard theorem, provides a necessary condition for linear instability of inviscid, stably stratified parallel shear flows. Specifically, if the gradient Richardson number, $Ri_g(z) = N^2/(d_z \overline{u})^2$, satisfies $Ri_g(z) > 1/4$ everywhere in the domain, then the flow is linearly stable. Conversely, if $Ri_g < 1/4$ in some region, instability is possible, though not guaranteed. This criterion has been successfully applied to stratified shear flows by \citet[][]{smith2021,olsthoorn2023}.  All studies confirmed that, for $Ri_g < 1/4$, shear instabilities lead to turbulent mixing, while for $Ri_g \gtrsim 1/4$ the flow tends to remain stable or only weakly turbulent. In the case of stratified Kolmogorov flow, \citet[][]{balmforth2002} performed a detailed linear stability analysis, identifying regions in parameter space where shear instabilities develop. They found that the marginal stability curve of the problem extends up to a maximum value of $Ri_g$ given by $Ri_g = 1/4$. They also showed that for sufficiently strong shear (i.e., small $Ri_g$), the flow becomes unstable to Kelvin–Helmholtz-like perturbations. The non-linear saturation of such instabilities typically leads to the generation of turbulence and the emergence of layered structures \citep[][]{balmforth2005}.

\subsection{Reynolds decomposition}
\label{subsec:rey}

The presence of a mean flow in the turbulent state suggests the use of a Reynolds decomposition of the velocity, pressure and density perturbation fields. The mean flow is in the same direction as the forcing (the $x$-direction) and, similarly to $\bm{f}(z)$, is horizontally invariant, depending only on the vertical direction $z$. This suggests a Reynolds decomposition based on horizontal averaging of the flow quantities. Hence, the velocity field is decomposed as $\bm{u} = \overline{\bm{u}} + \bm{u}'$, where $\overline{\bm{u}} (z,t)$ is the (horizontally averaged) mean flow and $\bm{u}'(\bm{x},t)$ are the fluctuations. The same decomposition can be applied to the pressure field $p$ and to the density perturbation $\theta$. 

Applying the Reynolds decomposition and averaging over horizontal directions $x,y$ the Eqs.~\eqref{eq1}--\eqref{eq2}, one obtains the equations for the mean flow:
\begin{eqnarray}
&& \partial_t \overline{u} + \partial_z \overline{u' w'} = \nu \partial_z^2 \overline{u} + F\cos(Kz), \label{M1} \\[0.2cm]
&& \partial_t \overline{v} + \partial_z \overline{v' w'} = \nu \partial_z^2 \overline{v}, \label{M2} \\[0.2cm]
&& \partial_z \overline{w'^2} = -\partial_z \overline{p} - N^2 \overline{\theta}, \label{M3} \\[0.2cm]
&& \partial_t \overline{\theta} + \partial_z \overline{\theta' w'} = \kappa \partial_z^2 \overline{\theta}. \label{M4} 
\end{eqnarray}

\noindent where incompressibility of both mean flow and fluctuations has been used, i.e. $\bm{\nabla}\cdot\overline{\bm{u}} = 0$ and $\bm{\nabla}\cdot\bm{u}' = 0$. In deriving the above equations, all terms containing horizontal derivatives vanish upon horizontal averaging, because of periodic boundary conditions. As a consequence, only the \(z\)-derivatives survive in the averaged equations. The mean incompressibility condition reduces to $\partial_z \overline{w}=0$, which implies $\overline{w}=\text{const}$.  As this constant can always be absorbed by a change of reference frame, we set $\overline{w}=0$ without loss of generality. Stationarity has not been assumed but would result in the cancellation of all remaining time derivatives in the above equations.

Let us consider the above equations and their consequences for our system. Equation~\eqref{M2} shows that a mean flow in the $y$-direction (the "spanwise" direction) can be generated starting from $\overline{v} = 0$, only under the action of the Reynolds stress $\partial_z \overline{v' w'}$. Similarly, equation~\eqref{M4} shows that a mean component of $\theta$ can only be generated through a non-zero $\partial_z \overline{\theta' w'}$, which represents the divergence of the turbulent (vertical) buoyancy flux. As we will see, both $\overline{v}$ and $\overline{\theta}$ are formed in our DNS and so they must have been fed by the turbulent fluctuations, which means that in this system energy is extracted from the turbulence to create a secondary mean flow and to alter the mean density profile. On the other hand, the primary mean flow, $\overline{u}$, feeds the turbulent fluctuations through the Reynolds stress in equation~\eqref{M1}, as expected for shear-generated turbulence. Furthermore, equation~\eqref{M3} shows that in this system, departure from mean hydrostatic balance is only due to the vertical Reynolds stress. 

Let us now consider the $x$-momentum equation, \eqref{M1}, in more detail. This equation firstly shows that $\overline{u}$ is maintained thanks to the forcing term on the RHS. In steady state, we expect a balance to exist between the Reynolds stress and this forcing term. Indeed, at high Reynolds number, the remaining viscous term should be small. This will be confirmed by DNS, which shows that $\overline{u}$ is almost two orders of magnitude smaller than the laminar amplitude $U_0$, which together with the fact that its variation in the $z$-direction is still controlled by the forcing wavenumber $K$, means that $\partial_z^2 \overline{u}$ is now much smaller than in the laminar case and so cannot balance the forcing. We can therefore write  $\partial_z \overline{u' w'} \approx  F \cos(Kz)$ for a statistically stationary state. A simple scaling assuming $\partial_z \sim K$ and $u' \sim w'$, gives $u' \sim \sqrt{F/K}$. This scaling implies that if the turbulent fluctuations are close to isotropic, then they will be controlled by the amplitude of the forcing and by $K$ and will be independent of the stratification $N$ or of the viscosity $\nu$. We will test this scaling against DNS results in the remaining sections. The velocity scale $U_F = \sqrt{F/K}$ was first introduced by \citet{garaud2016}, as a velocity scale for stratified Kolmogorow flow in a fully turbulent state. We will call $U_F$ the forcing velocity scale.

Equations~\eqref{M1}--\eqref{M4} can be subtracted from the full equations, \eqref{eq1}--\eqref{eq2}, to obtain equations for the turbulent fluctuations $u'$, $v'$, $w'$ and $\theta'$. These equations will be given in the appendix \ref{app:a}, as well as the equations for the mean flow kinetic and potential energy and for the turbulent kinetic and potential energy.

\section{Numerical simulations}
\label{sec:sim}

\begin{table}
\centering
\begin{tabularx}{\textwidth}{*{1}{X}p{1.25cm}*{11}{X}}
$M_x$ & $M_y \times M_z$ & $N^2$ & $Fr_F$ & $Re_F$ & $\ell_O$ & $\ell_B$ & $T$ & $\Gamma$ &  $Re$ & $Fr_t$ & $Re_b$ & {\rm Reg.} \\
\hline
$256$ & $256 \times 256$ & $0.00$ & $\infty$ & $89$ & $\infty$ & $\infty$ & $35.8$ & $0.00$ &  $229$ & $\infty$ & $\infty$ & S \\
$256$ & $256 \times 256$ & $0.005$ & $1.26$ & $89$ & $1.84$ & $1.57$ & $34.6$ & $0.13$  & $306$ & $1.21$ & $212$ & S\\
$256$ & $256 \times 256$ & $0.01$ & $0.89$ & $89$ & $1.21$ & $1.12$ & $36.0$ & $0.14$  & $382$ & $1.01$ & $128$ & S\\
$256$ & $256 \times 256$ & $0.02$ & $0.63$ & $89$ & $0.79$ & $0.80$ & $38.6$ & $0.15$  & $474$ & $0.84$ & $76$ & S\\
$256$ & $256 \times 256$ & $0.05$ & $0.40$ & $89$ & $0.45$ & $0.52$ & $45.5$ & $0.16$  & $665$ & $0.67$ & $40$ & S\\
$256$ & $256 \times 256$ & $0.07$ & $0.34$ & $89$ & $0.37$ & $0.44$ & $49.4$ & $0.16$  & $763$ & $0.62$ & $32$ & S\\
$256$ & $256 \times 256$ & $0.1$ & $0.28$ & $89$ & $-$ & $-$ & $-$ & $-$  & $-$ & $-$ & $-$ & I\\
\hline
$512$ & $512 \times 512$& $0.0$ & $\infty$ & $358$ & $\infty$ & $\infty$ & $9.3$ & $0.00$ & $993$ & $\infty$ & $\infty$ & S\\
$512$ & $512 \times 512$& $0.1$ & $1.13$ & $358$ & $1.74$ & $1.47$ & $9.4$ & $0.10$ & $1554$ & $1.27$ & $864$ & S\\
$512$ & $512 \times 512$& $0.2$ & $0.80$ & $358$ & $1.13$ & $1.04$ & $10.0$ & $0.12$  & $1913$ & $1.04$ & $509$ & S\\
$512$ & $512 \times 512$& $0.5$ & $0.51$ & $358$ & $0.63$ & $0.67$ & $11.2$ & $0.14$  & $2481$ & $0.79$ & $250$ & S\\
$512$ & $512 \times 512$& $1.0$ & $0.36$ & $358$ & $0.41$ & $0.48$ & $12.4$ & $0.15$  & $3036$ & $0.64$ & $145$ & S\\
$512$ & $512 \times 512$& $2.0$ & $0.25$ & $358$ & $0.27$ & $0.34$ & $14.5$ & $0.16$  & $3839$ & $0.53$ & $86$ & S\\
$512$ & $512 \times 512$& $4.0$ & $0.18$ & $358$ & $0.18$ & $0.24$ &  $17.5$ & $0.16$ & $5149$  & $0.45$ & $54$ & S\\
$512$ & $512 \times 512$& $5.0$ & $0.16$ & $358$ & $0.16$ & $0.22$ & $18.9$ & $0.16$  & $5739$ & $0.44$ & $47$ & S\\
\hline
$1024$ & $256 \times 256$ & $0.00$ & $\infty$ & $89$ & $\infty$ & $\infty$ & $150$ & $0.00$ & $26$ & $\infty$ & $\infty$ & S\\
$1024$ & $256 \times 256$ & $0.01$ & $0.89$ & $89$ & $0.75$ & $0.20$ & $60$ & $0.25$  & $177$ & $0.29$ & $56$ & S\\
$1024$ & $256 \times 256$ & $0.25$ & $0.18$ & $89$ & $-$ & $-$ & $-$ & $-$  & $-$ & $-$ & $-$ & I\\
$1024$ & $256 \times 256$ & $1.00$ & $0.09$ & $89$ & $-$ & $-$ & $-$ & $-$  & $-$ & $-$ & $-$ & I\\
\hline
\end{tabularx}
\caption{Simulation details: longitudinal grid resolution $M_x$, transversal grid resolutions $M_y \times M_z$, Brunt-V\"ais\"al\"a frequency $N^2$, forcing-scale Froude number $Fr_F=(FK)^{1/2}/N$, forcing-scale Reynolds number $Re_F=F^{1/2}/(\nu K^{3/2})$, Ozmidov scale $\ell_O=(\epsilon/N^{3})^{1/2}$, buoyancy scale $\ell_B=|\bm{u}'|_{\rm rms}/N$, eddy turnover time $T=\langle \bm{u}^2\rangle/(2\epsilon)$, mixing coefficient $\Gamma=\epsilon_p/\epsilon$, Reynolds number $Re=U/(\nu K)$ based on mean flow amplitude, turbulent Froude number $Fr_t=\epsilon/(N |\bm{u}'|_{\rm rms}^2)$, buoyancy Reynolds number $Re_b=\epsilon/(\nu N^2)$, regimes (Reg.), S: stationary, I: temporally intermittent. For intermittent runs, output parameters are not listed as they undergo large temporal oscillations. Fixed parameters: molecular viscosity and diffusivity $\nu=\kappa=10^{-3}$ (i.e. Prandtl number $Pr=1$), forcing wavenumber $K=1$, forcing amplitude $F=0.008$ for $M_x=256$ and $M_x=1024$ (i.e. $Re_F=89$), $F=0.128$ for $M=512$ (i.e. $Re_F=358$).}
\label{tab:param}
\end{table}

We solve Eqs.~\eqref{eq1}–\eqref{eq2} by means of DNS in a triply periodic domain, using a pseudo-spectral solver with dealiasing based on the $2/3$ rule and a second-order Runge–Kutta time integration scheme. The spatial grid resolution along each cartesian direction is defined as $(M_x, M_y, M_z)$ with sizes $(L_x, L_y, L_z)$. The main set of simulations is configured on a cubic domain $M=256$ and $M=512$ and domain size $L=2\pi$ (i.e. with $M = M_x = M_y = M_z$ and $L = L_x = L_y = L_z$). We ensure adequate resolution of small scales via the criterion $k_{\max} \, \eta \ge 2$, where $k_{\max} = M/3$ is the maximum resolved wavenumber and $\eta = (\nu^3 / \epsilon)^{1/4}$ is the Kolmogorov scale. 

We explore the effect of stable stratification by increasing the buoyancy frequency $N$. Starting from the unstratified case $N=0$, which reproduces the benchmark turbulent regime reported from previous studies \citep{musacchio2014}, we pushed $N$ up to its maximum value, beyond which a steady turbulent state is no longer sustainable and the flow becomes temporally intermittent. In the unstratified configuration ($N=0$), we begin from the laminar Kolmogorov flow and let the system evolve until a turbulent stationary state is established, discarding the initial transient dynamics. The resulting turbulent steady state is then used as the initial condition for the stratified runs. For moderate stratification, runs are directly initialized from that turbulent reference field and integrated until a new statistically stationary state is reached. For stronger stratification, we employ an incremental approach to mitigate excessively long transients: we progressively increase $N$, using the end state of a run at intermediate $N$ as the initial condition for the next, more stratified case. 

Simulation parameters are summarized in Table \ref{tab:param}. For each run, "input" and "output" parameters are given. The input parameters are the viscosity $\nu$, the diffusivity $\kappa$, the Brunt-V\"ais\"al\"a frequency $N^2$, the forcing amplitude $F$ and the forcing wavenumber $K$, in addition to the grid resolution. Together, these physical input parameters allow the definition of forcing-scale Froude and Reynolds numbers, $Fr_F = (FK)^{1/2}/N$ and $Re_F = F^{1/2}/(\nu K^{3/2})$, which are therefore also input parameters for each simulation (more details on $Fr_F$ and $Re_F$ are given in \S\ref{subsec:peak}). On the other hand, parameters involving flow quantities that are unknown at the start of each run, such as mean flow amplitude $U$ and kinetic energy dissipation $\epsilon$, are output parameters, and a number of these are also given in Table~\ref{tab:param}.

At strong stratification, turbulence remains only in bursts separated by quiescent phases, i.e. we enter a temporally intermittent regime. This is as opposed to all other runs with weaker stratification, for which a statistically stationary state with sustained turbulence is reached. For $M = 256$ we find that turbulence is no longer sustained at $N^2 = 0.1$, at which the flow becomes intermittent (see Table~\ref{tab:param}). At $M = 512$ a similar threshold is approached but never crossed, due to considerably slower evolution and prohibitive computational costs of achieving statistical convergence. Therefore, to further capture the dynamics of the intermittent regime, we employ an elongated domain with $M_x = 1024$, $M_y = M_z = 256$ with dimensions $L_x = 8\pi$, $L_y = L_z = 2\pi$. The use of elongated boxes allows us to adopt a higher nominal forcing wavenumber (given by $L_x/L_z = 4$) without sacrificing scale separation between forcing and dissipation, which in turn shortens the time of a cyclic oscillation of energetic phases (see Fig.~\ref{fig:ts}).
and accelerates the temporal dynamics. This is advantageous for observing the system passing through several cycles that would evolve too slowly in cubic domains. To illustrate the impact of domain geometry, we compare two simulations at $N^2 = 0.01$: one in a cubic domain ($256^3$) and one in the elongated domain ($1024 \times 256^2$). Both runs share the same forcing-scale parameters ($Fr_F = 0.89$, $Re_F = 89$), but their output parameters differ substantially (see Table~\ref{tab:param}). In particular, we can deduce that the elongated geometry helps access lower Froude numbers and lower buoyancy Reynolds numbers, effectively enabling exploration of stronger stratification. However, we caution that the dynamics of the flow in cubic and elongated domains may differ, which means that the critical $N$ at which transition to an intermittent regime occurs may also differ. Therefore, comparisons between these two geometries should be made with care.

This study primarily investigates the turbulent regime in the cubic domains with $M=256$ and $M=512$. Additional runs in elongated domains were performed only to provide a more complete picture of the transition to an intermittent regime. For cubic cases, after discarding transients, we gather statistics over a time window of $250$ eddy-turnover times, recording $180$ instantaneous fields and the corresponding vertical profiles to ensure statistical convergence. In the following section, we present results from the simulations, all of which have been time-averaged over these $180$ steady-state snapshots in addition to being spatially averaged. In particular, vertical profiles $\overline{u}(z)$, $\overline{v}(z)$, etc., are obtained by first performing a horizontal average and then a time average of $u$, $v$, etc. As a result, the symbol $\overline{[\cdot]}$, which we used in \S\ref{sec:model} to indicate horizontal averaging only, will from now on denote both horizontal and time averaging. We also construct global quantities that provide a single scalar value for each simulation, as shown, for example, in Table~\ref{tab:param}. Global quantities are formed via volume and time averages of the flow field. For example, the dissipation is defined as $\epsilon = \nu \langle | \bm{\nabla}\bm{u}|^2 \rangle$ where $\langle \cdot \rangle$ indicates a volume average over the entire numerical domain, combined with a time average over the $180$ steady-state snapshots.

\section{Results}
\label{sec:results}

\subsection{Flow-field visualizations}
\label{subsec:field}

\begin{figure}
\centering
\includegraphics[width=0.99\textwidth]{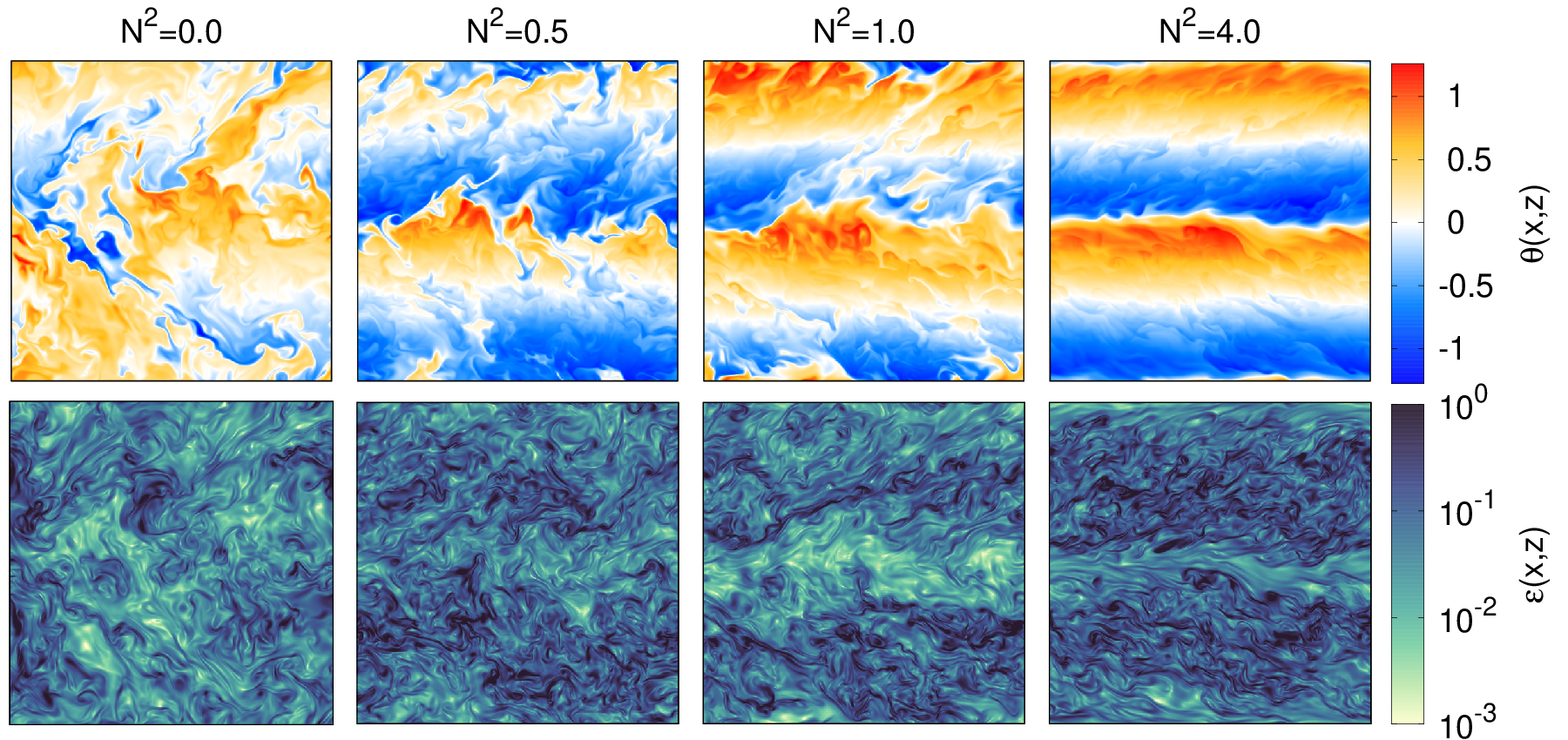}
\caption{Section visualisations of density perturbation $\theta$ (upper row) and kinetic energy dissipation rate $\varepsilon$ (lower row) in the plane $(x,z)$ and $y=L/2$. Resolution $M=512$.}
\label{fig:sect_te_film}
\end{figure}

In Figure~\ref{fig:sect_te_film} representative vertical sections of the scalar field $\theta(x,z)$ and of the local kinetic energy dissipation rate $\varepsilon(x,z)$ at fixed $y=L/2$ are presented, taken from simulations with $M=512$ with increasing levels of stratification, spanning from $N^2=0$ to $N^2=4$. In the absence of stratification ($N^2=0$), the scalar field $\theta$ exhibits fully developed small-scale turbulence with no apparent vertical organization. Structures oriented at 45$^\circ$ to the horizontal are observed, which are probably the result of turbulent production via shear. As stratification increases, the turbulence visible in the $\theta$-field --- $\theta$ can now be interpreted as a density perturbation --- appears less vigorous and less three-dimensional (even though the dissipation $\epsilon$ shown in the lower panels of Figure~\ref{fig:sect_te_film} does not decrease as $N^2$ is increased). The turbulent structures visible in the upper and lower half of the domain maintain an inclination to the horizontal but this is greatly reduced from 45$^\circ$, which is most likely due to the increased restoring force of buoyancy at higher $N$, reducing vertical motions. At the same time, as $N^2$ is increased, the density perturbation field becomes more and more organized along the vertical. There appears to be a mode-2 structure in the vertical direction, with two wavelengths of low-density-perturbation/high-density-perturbation over the vertical extent of the domain.

Turning to the vertical sections of kinetic energy dissipation rate $\varepsilon = \nu |\bm{\nabla}\bm{u}|^2$ in Figure~\ref{fig:sect_te_film}, they have a very different appearance compared to the $\theta(x,z)$ sections. The $\varepsilon$ field is much less structured than $\theta$ across all stratification strengths and looks rather homogeneous. The only exception is the fact that for the higher stratification levels, $N^2 = 1$ and $N^2=4$, there appear to be specific regions, close to $z/L = 0.5$ and $z/L=1$, where $\varepsilon$ is significantly lower than elsewhere in the domain. For $N^2=1$, these regions appear highly convoluted with large vertical excursions, while for $N^2=4$, the central and upper regions have become thinner and more clearly organized. Indeed, at $N^2 =4 $, the existence of two separate dynamical regions is most clearly displayed: there are two broad regions of high dissipation and 3D turbulence, one in the upper half of the domain and one in the lower half, separated by two thinner regions at $z/L = (0.5, 1)$, where flow structures are quasi-horizontal and the dissipation is up to 3 orders of magnitude lower (indicating, perhaps, relaminarization).   

\subsection{Mean velocity and density profiles}
\label{subsec:meanflow}

\begin{figure}
\centering
\includegraphics[width=0.99\textwidth]{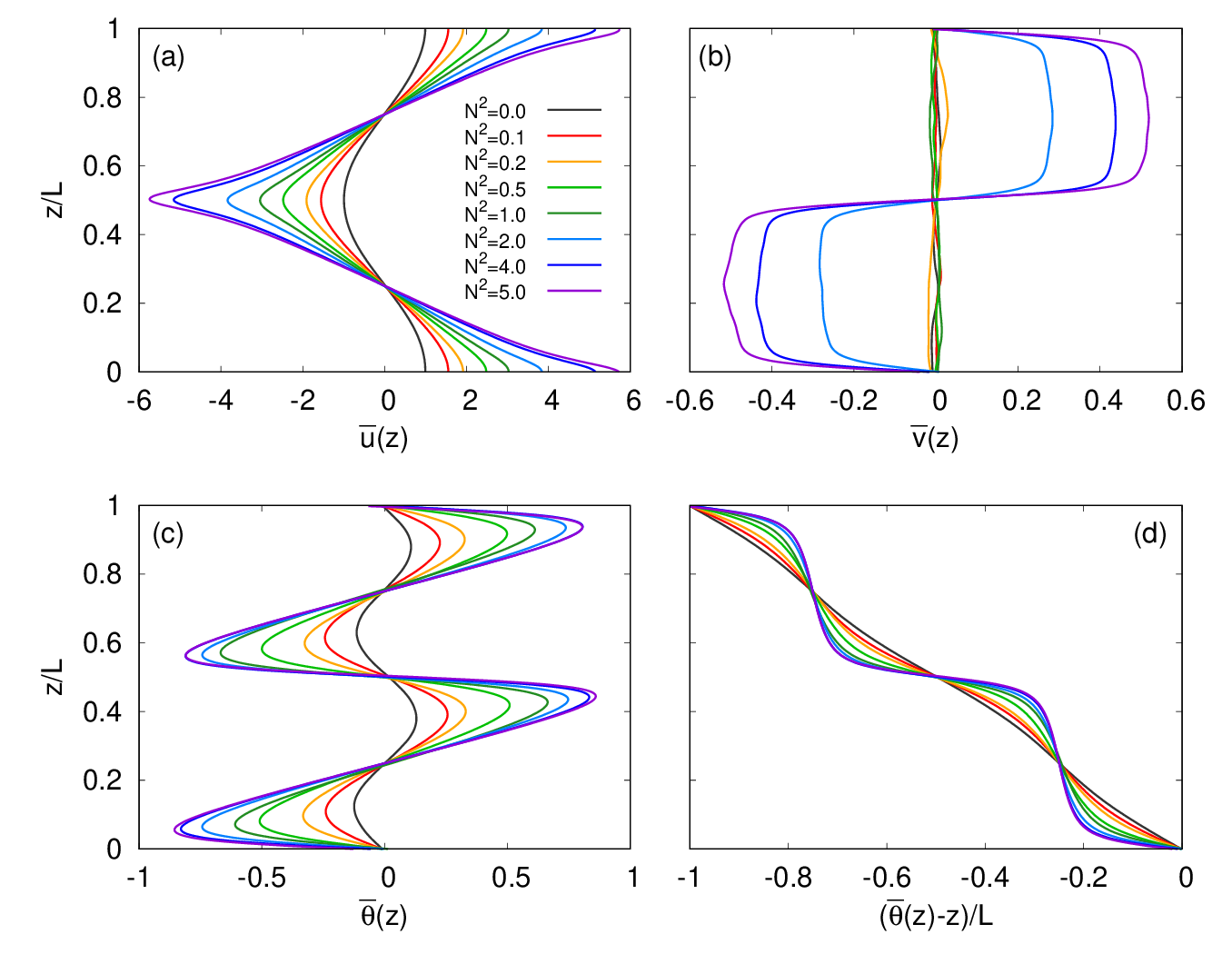}
\caption{Vertical profiles of mean flow: longitudinal velocity $\overline{u}(z)$ (upper left panel), spanwise velocity $\overline{v}(z)$ (upper right), density perturbation $\overline{\theta}(z)$ (lower left) and total density $\overline{\theta}(z)-z$ (lower right) across the runs at resolution $M=512$.}
\label{fig:profs}
\end{figure}

We begin our analysis by examining the vertical profiles of the mean flow that emerges during the statistically stationary phase of the simulations. As shown in Figure~\ref{fig:profs}(a), the primary mean flow, $\overline{u}(z)$, undergoes a marked transformation with increasing stratification. In the non-stratified regime ($N=0$), the velocity profile closely follows a monochromatic function of the form \( U \cos(K z) \), characteristic of the classical Kolmogorov flow (see, e.g., \citealt{musacchio2014}). As stratification increases (i.e. as \( N \) increases), the mean flow amplitude grows and the profile progressively distorts into a sawtooth-like shape, exhibiting two oppositely sheared regions separated by sharp transitions (as observed also by \citealt{garaud2016}). In the simulations with the highest stratification, the mean profile is further distorted with the appearance of additional inflection points creating "peaky" maxima and minima. 

At the same time, as shown in Figure~\ref{fig:profs}(b), the three simulations with highest $N$ present a mean spanwise flow $\overline{v}(z)$ of non-negligible magnitude, even though this remains one order of magnitude smaller than the longitudinal velocity $\overline{u}(z)$. The mean spanwise flow has a mode-1 structure, similarly to the mean longitudinal flow and to the forcing, but is 90$^\circ$ out of phase compared to $\overline{u}(z)$ and $\bm{f}$. In the upper half of the domain, $\overline{v}(z)$ presents (approximately) a positive plateau, while in the lower half, it presents a negative plateau, with strong gradients at $z=L/2$ and $z=L$. Because of this non-zero $\overline{v}(z)$ in the most stratified runs, the mean flow $\overline{\bm{u}}(z) = (\overline{u}(z),\overline{v}(z), 0)$ is not strictly a parallel shear flow in these runs. Given that the magnitude of $\overline{v}(z)$ is an order of magnitude smaller than $\overline{u}(z)$, this effect remains small and the flow is still close to a parallel shear flow. Why such a non-zero $\overline{v}(z)$ develops at high $N$ is not clear but equation~\eqref{M2} shows that this mean flow can grow from $\overline{v}=0$ only through the term $\partial_z \overline{v' w'}$, hence via some organization of the fluctuations $v'$ and $w'$. So it appears that the primary mean flow, $\overline{u}(z)$, creates turbulence via shear production and that this turbulence then returns some of its energy to feed a secondary mean flow, $\overline{v}(z)$. 

We now turn to the mean density profiles, presented in the lower panels of Figure~\ref{fig:profs}. Vertical profiles of $\overline{\theta}(z)$ and of $\overline{\theta}(z)-z$ are given. Note that in both cases a further multiplication by $N^2$ is required to get, respectively, the mean density perturbation $\overline{\rho}' g/ \rho_0$ and the total density $(\overline{\rho}-\rho_0) g/\rho_0$ (as a result the actual density profiles are much more spread out and would not all fit on the same graph). The mode-2 structure visible in the visualizations of $\theta(x,z)$ in Figure~\ref{fig:sect_te_film} is reproduced by the mean density perturbation $\overline{\theta}(z)$. Similarly to $\overline{u}(z)$, $\overline{\theta}(z)$ also increases with increasing $N$ and loses its sinusoidal dependence on $z$. Contrary to $\overline{u}(z)$ though, $\overline{\theta}(z)$ actually becomes asymmetric and its peaks are displaced, getting closer and closer to $z/L =(0.5,\ 1)$ as the stratification gets stronger. This behaviour results in the formation of a layer-interface structure in the total density profile as shown by the graph of $\overline{\theta}(z)-z$. Strongly stable interfaces with total density gradient $\ge\gamma$, the background density gradient, develop around $z/L =(0.5,\ 1)$, while weakly stable layers with total density gradient $<\gamma$ develop on either side of each interface and take up most of the vertical extent of the simulation. Such layer-interface structure, sometimes called a \textit{staircase profile}, is a ubiquitous feature of stratified turbulent flows, from toy models \citep{balmforth1998,ponetti2018}, to laboratory experiments \citep{park1994,holford1999}, numerical simulations \citep{radko2007,kimura2016,maffioli2019,kimura2024} and ocean measurements \citep{desaubies1981,pinkel1991,gregg2018}.

Comparing the profiles of $\overline{u}(z)$ and of $\overline{\theta}(z)-z$ in Figure~\ref{fig:profs}, one can see that the interfaces in the density profile are formed at the locations of the maximum and minimum of $\overline{u}(z)$. These extrema of $\overline{u}(z)$ are also the locations where the mean shear is zero, since $d_z \overline{u}=0$ there. Conversely, the weakly stratified layers correspond to the shear layers of $\overline{u}(z)$ in which $|d_z \overline{u}(z)|$ is maximum. Since the background stratification is uniform, the interface and layer positions are not externally imposed but result from the flow dynamics. A possible explanation is that regions of non-zero shear are susceptible to shear instabilities and mixing, which tend to homogenize the density and reduce local stratification, thus creating a layer. In contrast, regions of weak or vanishing shear do not undergo shear instability and act as preferred sites for the formation of density interfaces, whose strong density gradient further increases the stability of that location in space. The presence of interfaces in between the layers is indeed necessary to ensure that the overall density change over the height of the domain remains unchanged (as it must since this is externally imposed). If this argument is true it means that the density profile is essentially slaved to the mean velocity profile and to its spatial distribution of vertical shear. It explains that the wavelength $L/2$ observed for the layer-interface structure of $\overline{\theta}(z)-z$ is set by the wavelength of a high-shear/low-shear sequence in $\overline{u}(z)$, which is also $L/2$.

\subsection{Variation of mean flow amplitude}
\label{subsec:peak}

We quantitatively investigate the variation of mean flow across our DNS dataset. We begin by focusing on the increase of the mean flow $\overline{u}(z)$ with increasing stratification, as highlighted by the profiles in Figure~\ref{fig:profs}. We study this increase in velocity by considering the peak amplitude of the velocity profile, defined as $U = 0.5 \bigl(\max\{\overline{u}(z)\} - \min\{\overline{u}(z)\}\bigr)$. We will refer to $U$ as the amplitude of the mean flow, even though it is strictly the mean flow amplitude only when $\overline{u}(z)$ is sinusoidal and when $\overline{v}(z) = 0$. To understand how $U$ varies with the imposed flow parameters, we consider a simple dimensional analysis. Given that $Pr=1$ and so $\nu = \kappa$, the variables that $U$ could depend on are:

\begin{equation}
U = U(F,K,N,\nu)
\end{equation}

\noindent
meaning that there are a total of 5 variables for 2 dimensions (length and time) so that Buckingham's Pi theorem tells us that we can form $5-2=3$ dimensionless groups. Considering our simulations are turbulent, we choose to nondimensionalise $U$ by the forcing velocity scale $U_F = \sqrt{F/K}$ and not by the laminar velocity amplitude $U_0$. For the two remaining dimensionless groups, we choose to form Froude and Reynolds numbers based on $U_F$ and $K$, closely following the choice of \citealt{garaud2016} (we use their symbol naming convention and call them $Fr_F$ and $Re_F$),

\begin{equation}
\frac{U}{U_F} = \frac{UK^{1/2}}{F^{1/2}}, \hspace{20pt} 
Fr_F = \frac{U_F K}{N} = \frac{F^{1/2}K^{1/2}}{N}, \hspace{20pt} 
Re_F = \frac{U_F}{\nu K} = \frac{F^{1/2}}{\nu K^{3/2}}.
\end{equation}

\begin{figure}
\centering
\includegraphics[width=0.99\textwidth]{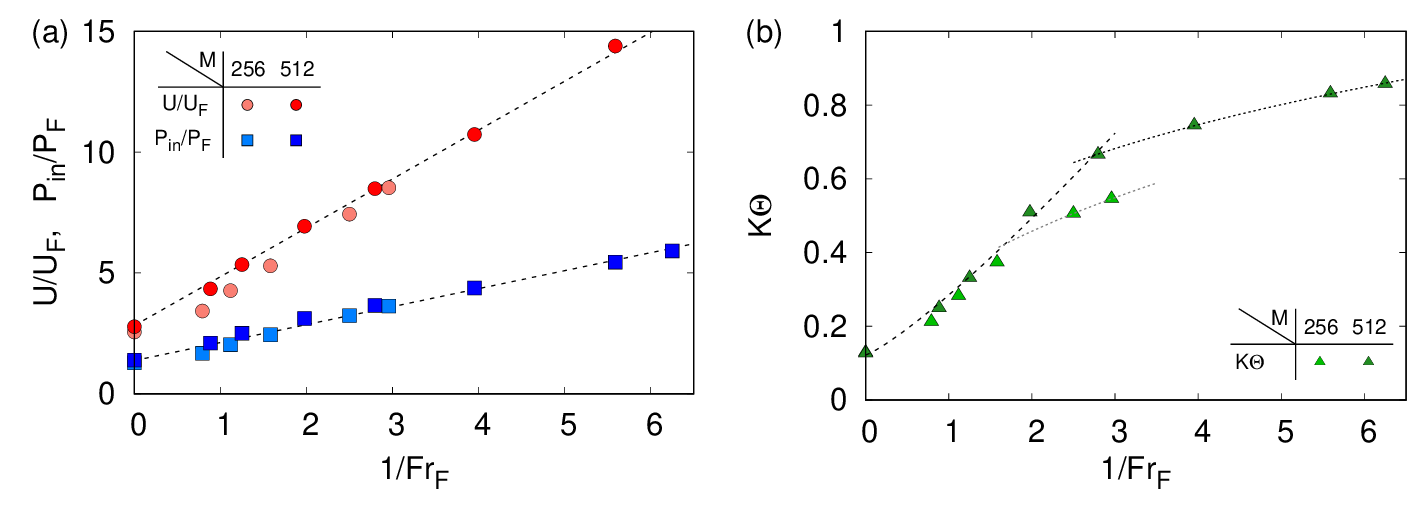}
\caption{Left panel: Evolution of the mean-flow amplitude $U$ (red circles) and input power $P_{\rm in}$ (blue squares) as a function of $1/Fr_F$ for all DNS runs. The data are shown in dimensionless form, with linear fits included. Right panel: Density amplitude $\Theta$ (normalized by $K = 1$) as a function of $1/Fr_F$. At low $1/Fr_F$ (weak stratification), $\Theta$ follows a power law, $\Theta \propto Fr_F^\beta$, with $\beta = 1.2$ (dashed black). At high $1/Fr_F$ (strong stratification), separate power-law curves are observed for the two resolutions: $M=512$ with $\beta = 0.3$ (solid black) and $M=256$ with $\beta = 0.4$ (dotted gray). Other fit parameters are omitted for clarity.}
\label{fig:peak}
\end{figure}

We can now look for the dependence of $U/U_F$ on $Fr_F$ and $Re_F$ in our DNS results. In Figure~\ref{fig:peak}(a) we show $U/U_F$ as a function of $1/Fr_F$, which shows a good collapse of all data points, for both 256$^3$ and $512^3$ runs, on a single curve which is close to linear. It may be that a weak $Re_F$-dependence remains, as evident from the fact that two sets of points are distinguishable from Figure~\ref{fig:peak}(a), one below the linear fit and one above it, roughly corresponding to results from the 256$^3$ runs on the one hand and from the $512^3$ runs on the other. We choose not to explore this Reynolds number dependence, considering that the linear fit in Figure~\ref{fig:peak}(a) is quite good, and from now on use the result that $U/U_F \approx a Fr_F^{-1} + b$, where $a$ and $b$ are constants. Given that $Fr_F = U_F K/N$, the mean flow amplitude can be written as $U \approx a N/K + b U_F$ and thus contains two contributions, one involving the stratification $N$ and one involving $U_F$ and hence the forcing amplitude $F$. The left panel of Figure~\ref{fig:peak} shows that at high stratification, i.e. high $Fr_F^{-1}$, the relative contribution of the term involving $N$ to $U$ becomes dominant over the contribution of the term containing $U_F$.

Turning to the consequences of this change in $U$ across the DNS runs, the external force is purely in the $x$-direction and this means that the input power $P_\text{in}= \langle \bm{u}\cdot\bm{f} \rangle = (1/2\pi) \int_0^{2\pi} \overline{u}(z) \ F \cos (Kz)\ dz$ should behave similarly to $\overline{u}$ and to $U$. The left panel of Figure~\ref{fig:peak} shows the variation of the nondimensionalised input power, $P_\text{in}/P_F$, as a function of $1/Fr_F$. As expected, $P/P_F$ also increases linearly with $Fr_F^{-1}$ and so has a similar behaviour to $U/U_F$. Specifically, we find $P/P_F = c Fr_F^{-1} + d$ with $c=0.74$ and $d=1.37$ is a good fit to the data. The input power is nondimensionalised using $P_F = U_F^3 K$, which could be thought of as a dissipation scaling based on $U_F$ and $K$. The input power therefore increases significantly as $N$ is increased throughout the different runs, as a direct consequence of the increase in $U$. 

We proceed to consider the variation of the mean density profile across the simulations. The density amplitude $\Theta$ is shown in the right panel of Figure~\ref{fig:peak}, as a function of $Fr_F^{-1}$. Following the definition of $U$, we define $\Theta = 0.5\bigl(\max\{\overline{\theta}(z)\} - \min\{\overline{\theta}(z)\}\bigr)$. Unlike $U$, however, $\Theta$ does not increase linearly with $Fr_F^{-1}$ across the simulations. Instead, two distinct regimes can be identified: i) a relatively fast, almost linear increase with $Fr_F^{-1}$, approximately following $\Theta \sim Fr_F^{-1.2}$, at weak stratification, and ii) a slower growth, roughly $\Theta \sim Fr_F^{-0.3}$, at stronger stratification. The transition between these regimes appears to occur for $Fr_F^{-1}$ in the range $[2,\;3]$ (i.e. $Fr_F \in [0.33,\;0.5]$) for both $M=256$ and $M=512$ simulations.

\subsection{Scaling and anisotropy of turbulent fluctuations}
\label{subsec:primes}

We compare our DNS runs with the non-stratified simulations of turbulent Kolmogorov flow of \citet{musacchio2014}, hereinafter denoted MB14, in which the forcing amplitude $F$ was varied to vary the Reynolds number of the simulations. In our DNS, $F$ is kept constant for a given resolution ($M=256$ or 512), while the Brunt-V\"{a}is\"{a}l\"{a} frequency $N$ is varied instead. This means that only two values of $Re_F$ are explored in our DNS, yet the mean flow amplitude $U$ varies continuously across the DNS dataset, as we have seen in the preceding section. This calls for an improved version of the Reynolds number, which should be representative of the mean flow, and a good candidate is $Re = U/\nu K$. Note that since $\nu$ and $K$ are constant across the DNS dataset, this is just a rescaling of $U$. Using the value of $Re$ across the simulations, we can compare against those of MB14.

\begin{figure}
\centering
\includegraphics[width=0.99\textwidth]{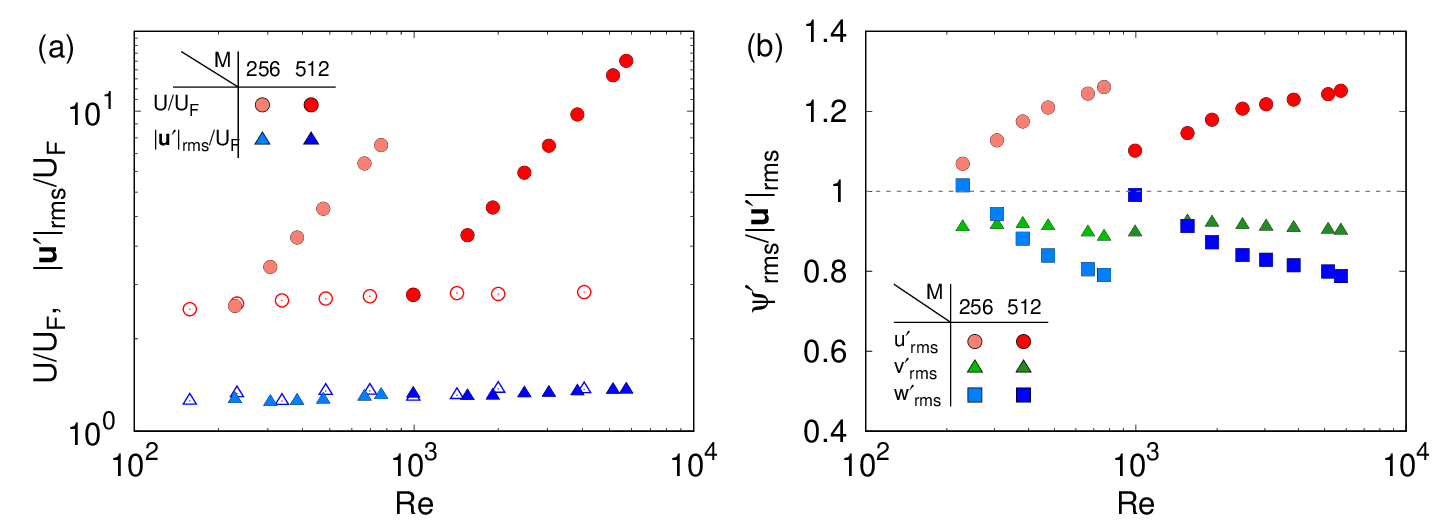}
\caption{Left panel: mean flow amplitude, $U/U_F$, and RMS turbulent velocity, $|\bm{u}'|_{\rm rms}/U_F$, versus Reynolds number, $Re=U/\nu K$. The empty symbols correspond to the data of \citet{musacchio2014}, which was converted to the present form using the values given in Table 1 of \citet{musacchio2014}. Right panel: RMS turbulent velocity components, normalized by $|\bm{u}'|_{\rm rms}$, versus $Re$. Isotropy of the turbulent velocity components would give a value of 1 for these normalized quantities.}
\label{fig:urms}
\end{figure}

For each resolution, we carried out a non-stratified run with $N^2=0$, which we can compare to the results of MB14. We can also assess how well the data of MB14 agree with the linear dependence of $U/U_F$ with $Fr_F^{-1}$, which was found in \S\ref{subsec:peak} to be a good fit to the present data. For $N=0$, this linear fit simply becomes $U/U_F = b = {\rm const}$ so that the mean flow amplitude should scale as $U_F$, $U \sim U_F$. Moreover, in \S\ref{subsec:rey}, we found that also the turbulent fluctuations should scale as $U_F$, for both stratified and non-stratified Kolmogorov flow. This prediction can be written as $|\bm{u}'|_{\rm rms} \sim U_F$, or, since isotropy was assumed, as $u'_{\rm rms} \sim v'_{\rm rms} \sim w'_{\rm rms} \sim U_F$, where 
$$
|\bm{u}'|_{\rm rms} = \sqrt{\dfrac{\langle|\bm{u}'|^2 \rangle}{3}},
\quad\mbox{and}\quad 
u'_{\rm rms} = \sqrt{\langle u'^2 \rangle},\quad
v'_{\rm rms} = \sqrt{\langle v'^2 \rangle},\quad 
w'_{\rm rms} = \sqrt{\langle w'^2 \rangle}.
$$

The left panel of Figure~\ref{fig:urms} shows the evolutions of $U/U_F$ and $|\bm{u}'|_{\rm rms}/U_F$ across the present DNS runs and the unstratified DNS runs of MB14, as a function of $Re$. Focusing on $U/U_F$, the first thing to note is that our present results are in excellent agreement with the results of MB14, as highlighted by the two data points corresponding to our runs with $N^2=0$ practically coinciding with two data points from MB14. The present data then departs from the data of MB14, as expected, as these data points correspond to the stratified runs with $N^2>0$. Moreover, the unstratified runs of MB14 and of the present dataset show that $U/U_F$ remains approximately constant over the entire range of $Re$ values. This confirms the unstratified scaling of the mean flow amplitude, $U \sim U_F$. 

We now turn to the evolution of $|\bm{u}'|_{\rm rms}/U_F$ with $Re$. In this case a good collapse of all data points is observed, comprising data from MB14 and from the present simulations, both stratified and unstratified. The data collapse on an approximately constant plateau, which confirms the scaling $|\bm{u}'|_{\rm rms} \sim U_F$ found in \S\ref{subsec:rey} and validates it for both stratified and unstratified Kolmogorov flow. The comparison of the present data with the dataset of MB14 has therefore confirmed that, in unstratified turbulent Kolmogorov flow, both the mean flow amplitude and the RMS turbulent fluctuations scale as $U\sim U_F$ and $|\bm{u}'|_{\rm rms}\sim U_F$. As far as the authors know, this result has not been previously reported for turbulent Kolmogorov flow, even though MB14 did report that mean flow amplitude and turbulent fluctuations were proportional to one another, $U \sim |\bm{u}|'_{\rm rms}$ (which is of course consistent with them both being proportional to $U_F$).

On the right panel of Figure~\ref{fig:urms}, the RMS turbulent velocity components, $u'_{\rm rms}$, $v'_{\rm rms}$, $w'_{\rm rms}$, normalized by $|\bm{u}'|_{\rm rms}$, are shown as a function of $Re$ for the present simulations only. The data fall within the range of values [0.7, 1.3], indicating some departure from the isotropic value of 1. These departures from isotropy are probably not large enough to invalidate the scaling arguments of \S\ref{subsec:rey} in which it was assumed that $u' \sim w'$, since, despite different trends with varying $Re$, $u'_{\rm rms}$ and $w'_{\rm rms}$ remain of the same order of magnitude. The departure from isotropy increases with increasing $Re$, starting from approximate isotropy at low Reynolds number (the lowest $Re$ at both resolutions corresponds to the runs with $N^2=0$) and going towards anisotropic conditions with an increase in $u'_{\rm rms}$ and a decrease in $w'_{\rm rms}$ as $Re$ increases. As seen before, the increase in $U$ and therefore in $Re$ in both sets of DNS runs, at $M=256$ and $M=512$, is a result of the increase in stratification values, so the departure from isotropy and, in particular, the damping of vertical velocity fluctuations, as shown by the decrease in $w'_{\rm rms}$, is a result of the increasing restoring force imparted by the stratification. The corresponding increase in $u'_{\rm rms}$, which appears anti-correlated to the decrease in $w'_{\rm rms}$, could be due to the fact that longitudinal and vertical velocity fluctuations are related by the fact that they are both directly fed by the shear, possibly through shear instabilities. Hence the concurrent evolution of $u'_{\rm rms}$ and $w'_{\rm rms}$ could reflect the fact that turbulent structures related to shear production are more and more horizontal as stratification is increased (as shown by the visualizations of Figure~\ref{fig:sect_te_film}). Indeed, we expect flattened structures with little inclination to the horizontal to have $u'>w'$. It is finally worth pointing out that the transverse velocity fluctuations, quantified by $v'_{\rm rms}/|\bm{u}'|_{\rm rms}$, show little to no variation across the DNS runs, remaining close to the isotropic value of unity.

\subsection{Richardson number profiles}
\label{subsec:rig}

\begin{figure}
\centering
\includegraphics[width=0.99\textwidth]{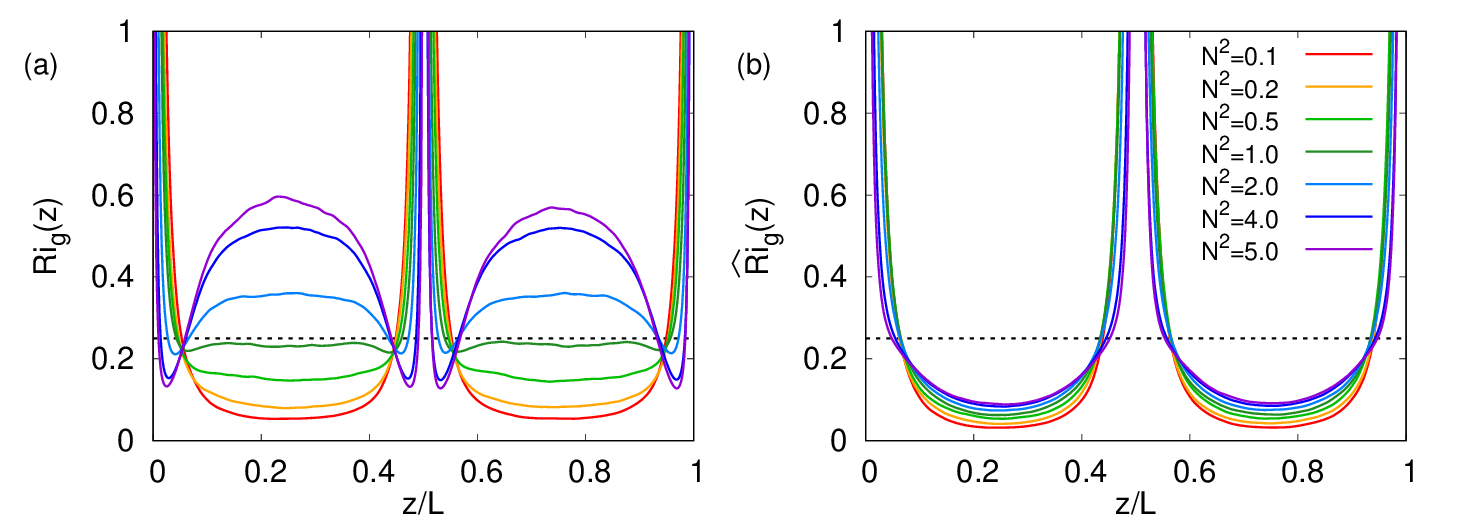}
\caption{Vertical profiles of the gradient Richardson number: ${Ri}_g(z) = N^2 / (d_z \overline{u})^2$ (left panel) and $\widehat{Ri}_g(z) = [\,N^2 (1 - d_z \overline{\theta})\,] / (d_z \overline{u})^2$ (right panel). The dashed line marks the critical value $Ri_g = 0.25$. Results are shown for simulations at resolution $M = 512$.}
\label{fig:ri}
\end{figure}

In \S~\ref{subsec:peak} we showed that the mean-flow amplitude increases approximately linearly with stratification, $U \propto N$. A possible rationalization of this result is that stronger stratification stabilizes the flow against shear instabilities; therefore, to maintain a level of instability sufficient to feed the turbulence and dissipate the input power, the mean shear must increase. Since the mean velocity profile is nearly linear in the two regions adjacent to $z=L/2$, the shear can be approximated as $d_z \overline{u} \approx UK$. At strong stratification (high $N$ and $Fr_F^{-1}$), where $U\sim N/K$, this scaling implies $d_z \overline{u} \sim UK \sim N$. As a result, the gradient Richardson number, introduced in \S~\ref{subsec:eqs}, satisfies $Ri_g = O(1)$ at high $N$, which is compatible with the development of shear instabilities. The classical criterion for shear instability is $Ri_g \leq 0.25$, but in turbulent flows it is commonly used in a weaker form $Ri_g \lesssim 1$.

To test these ideas, we analyse the mean vertical profiles of the gradient Richardson number using two definitions: the classical form ${Ri}_g(z) = N^2/(d_z\overline{u})^2$, and a modified form, $\widehat{Ri}_g(z) = N^2(1-d_z\overline{\theta})/(d_z\overline{u})^2$, which accounts for the modification of the local density gradient by the mean flow. 

Vertical profiles of $Ri_g$ and $\widehat{Ri}_g$ for the $M=512$ simulations are shown in Figure~\ref{fig:ri}. Both measures display similar behavior and reveal a two-region structure that closely follows the layer-interface organization of the mean density profile described in Figure~\ref{fig:profs}. Thus, the flow may be idealized as alternating wide bulk layers and thin interfaces, repeated twice due to the forcing with $K=1$.

The interface regions, around $z/L \in (0.5,1)$, exhibit $Ri_g$ and $\widehat{Ri}_g$ well above unity, indicating strong stability and suppressed shear instabilities, consistent with their role as barriers to vertical mixing. They are therefore "shielded" against potential shear instability that could lead to its depletion by mixing. The interiors of density layers occupy a broader region centered around $z/L = (0.25,\ 0.75)$, and they correspond to the layers on the mean density profile of Figure~\ref{fig:profs}d. Here the vertical shear $d_z \overline{u}$ is large, and both definitions of the gradient Richardson number take relatively low values, of order unity or smaller. 

The critical value of $Ri_g$ from linear stability theory, $Ri_g = 0.25$, has been added to the plots for reference. This value retains some dynamical significance even in turbulent conditions: all profiles intersect at this value and it seems to play the role of a threshold in the transition to strong stratification. This is consistent with the recent theoretical work by \citet{chini2022,shah2024} who showed in their multiscale analysis of stratified turbulence that the turbulent fluctuations obey quasilinear dynamics to leading order.

In the simulations, as stratification increases, a systematic trend emerges. In the bulk, the mean value of $Ri_g$ increases gradually with $N^2$, evolving from subcritical values ($Ri_g<1/4$) at weak stratification to slightly supercritical values ($Ri>1/4$) at the strongest stratification considered. At the same time, the interface develop an internal organization. An interfacial core, maintaining $Ri_g \gg 1$, becomes progressively thinner, while buffer layers appear on either side of the central core with $Ri_g<1/4$, indicating zones where shear can overcome stratification even while the central core remains strongly stable. 
The coexistence of a highly stable core and adjacent turbulent buffer zones is consistent with observations in strongly stratified shear flows, where a reduction of turbulence in the bulk often shifts shear production toward the interfaces, generating localised turbulent patches or Kelvin–Helmholtz-like activity. This behaviour is often referred to as turbulent "scouring" of the interface \citep{smith2021}.

Overall, the profiles depict a flow that self-organises into increasingly sharp and stable interfaces separated by bulk layers that remain marginally stable or weakly unstable. At strong stratification, part of the turbulent activity shifts from the bulk to thin interfacial buffer zones, while the central interfacial cores become progressively thinner and more stable. The continual occurrence of shear instabilities within these bulk and buffer regions supplies a sustained source of turbulence that maintains the layered mean-flow structure.

\subsection{Dimensionless parameters relevant for the turbulence}
\label{subsec:output-param}

Having characterized the mean flow, turbulent fluctuation magnitude, and Richardson numbers, we now examine the global dimensionless parameters that capture the balance between turbulence and stratification, and thus help characterize the dynamical regime. Here we focus on global quantities —one representative value for each run— rather than on vertical profiles or other, more local, diagnostics. Specifically, we consider the turbulent Froude number $Fr_t = \epsilon / (N|\bm{u}'|^2_{\rm rms})$ and the buoyancy Reynolds number $Re_b = \epsilon / (\nu N^2)$, computed for each simulation. The kinetic energy dissipation rate is defined as $\epsilon = \nu \langle |\bm{\nabla}\bm{u}|^2\rangle$, where $\bm{u}$ is the total velocity field. Although this definition includes both mean and fluctuating contributions, the dissipation is expected to be dominated by the small-scale gradients of the fluctuating velocity field, so that the use of total fields is fully justified.

In statistically steady conditions, the kinetic and potential energy dissipation rates balance the power input, $P_{in} = \epsilon + \epsilon_p$, where the potential energy dissipation rate is $\epsilon_p = \kappa N^2 \langle |\bm{\nabla}\theta|^2 \rangle$. Introducing the mixing coefficient $\Gamma = \epsilon_p / \epsilon$ and assuming it remains approximately constant and moderately small across our simulations, we obtain $P_{in} = \epsilon(1+\Gamma) \sim \epsilon$. This assumption is consistent with our data, as $\Gamma$ remains in the range $[0.1,\;0.16]$ in all cubic-domain simulations (see Table~\ref{tab:param}). This provides the following scaling prediction for $Re_b$,

\begin{equation}
Re_b = \frac{\epsilon}{\nu N^2} \sim \frac{P_\text{in}}{\nu N^2} = \frac{P_\text{in}}{P_F} \frac{P_F}{\nu N^2} \approx (cFr_F^{-1} +d) Re_F Fr_F^2 = (c + d Fr_F) Re_F Fr_F, 
\label{Reb_scaling}
\end{equation}

\noindent
where we have used the linear fit found in \S\ref{subsec:peak} for $P_{in}/P_F$ in terms of $Fr_F^{-1}$. The RHS of equation~\eqref{Reb_scaling} is fully expressed in terms of $Fr_F$ and $Re_F$, the forcing-scale Froude and Reynolds numbers, and includes a term $\propto Re_F Fr_F$ and a term $\propto Re_F Fr_F^2$. The latter term is reminiscent of the exact relation between $Re_b$ and $Fr_t$, $Re_b = Re_t Fr_t^2$, where $Re_t = |\bm{u}'|_{\rm rms}^4/(\nu\epsilon)$ is the turbulence Reynolds number \citep[see][]{brethouwer2007}.

To obtain a scaling for $Fr_t$ we use the additional fact that $|\bm{u}'|_{\rm rms} \sim U_F$, as predicted by the scaling analysis in \S\ref{subsec:rey} and as confirmed by the DNS results. Hence

\begin{equation}
Fr_t = \frac{\epsilon}{N|\bm{u}'|^2_{\rm rms}} \sim \frac{P_{in}}{N U_F^2} = \frac{P_{in}}{P_F}\frac{P_F}{N U_F^2} \approx  (cFr_F^{-1} +d) Fr_F = c + d Fr_F, \label{Frt_scaling} 
\end{equation}

\noindent
which also leads to a scaling result based only on forcing-scale parameters, this time based only on $Fr_F$. Equation~\ref{Frt_scaling} shows that $Fr_t$ and $Fr_F$ should be linearly related, a rather reassuring result, which confirms that $Fr_F$ is a Froude number relevant for the turbulent fluctuations. 

\begin{figure}
\centering
\includegraphics[width=0.99\textwidth]{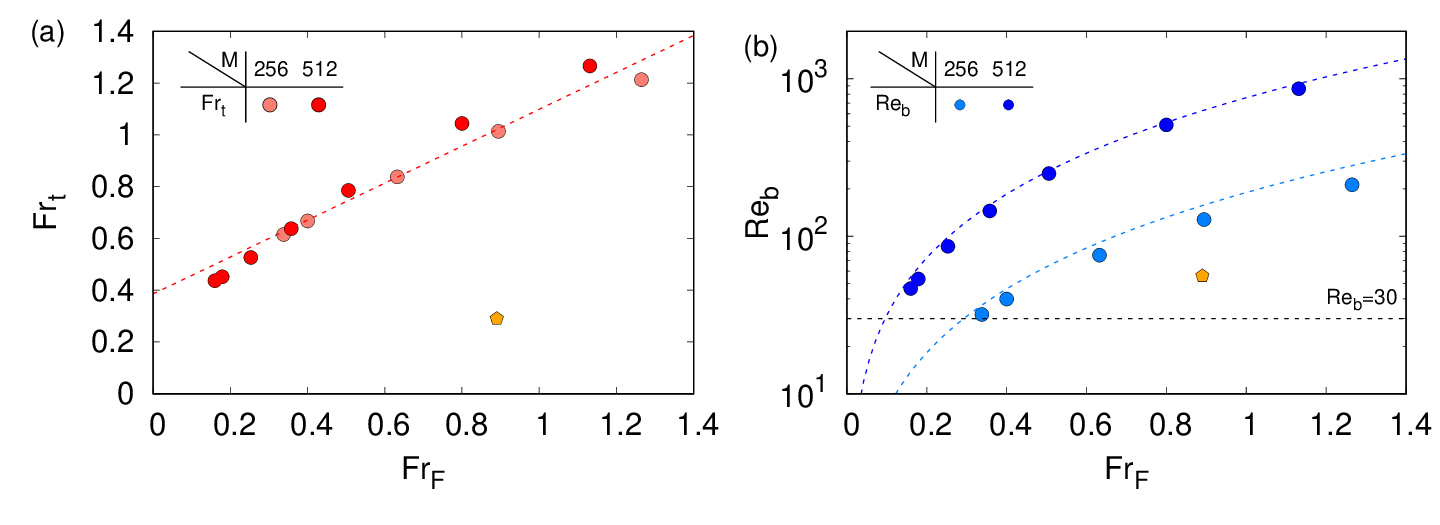}
\caption{Turbulent Froude number $Fr_t=\epsilon/(N|\bm{u}'|_{\rm rms}^2)$ (left panel) and buoyancy Reynolds number $Re_b = \epsilon/(\nu N^2)$ (right panel) as a function of the forcing-scale Froude number $Fr_F$. Dashed lines represent $Fr_t=\lambda(c+d Fr_F)$ and $Re_b = (c+d Fr_F) Re_F Fr_F$, with $c$ and $d$ taken from the linear fit of $P_{in}/P_F = f(Fr_F)$ given in Figure~\ref{fig:peak} ($c=0.74$, $d=1.37$) and $\lambda=0.52$. The orange pentagonal marker indicates the only statistically stationary simulation performed in the elongated domain.}
\label{fig:fr}
\end{figure}

Equations~\eqref{Reb_scaling}--\eqref{Frt_scaling} represent order of magnitude relations and additional order one constants may be needed for the RHS of these equations to be good approximations of $Re_b$ and $Fr_t$. This is the case for \eqref{Frt_scaling}, where an additional constant, $\lambda$, is needed to form the approximate relation, $Fr_t \approx \lambda (c + d Fr_F)$, which is plotted in Figure~\ref{fig:fr}. The data for both $M=256$ and $M=512$ is in good agreement with this relation, across all values of $Fr_F$. As for equation~\eqref{Reb_scaling}, it turns out that the RHS of \eqref{Reb_scaling} is already a relatively good approximation of $Re_b$, without the need for additional constants, so that $Re_b \approx Re_F (c Fr_F + d Fr_F^2)$ (this is essentially because the relatively low values of $\Gamma$ mean that $\epsilon \approx P_{in}$ in the simulations). This is demonstrated by the graph of $Re_b$ as a function of $Fr_F$ in Figure~\ref{fig:fr}. Note that two curves are shown in this graph because there are two different values of $Re_F$ within the DNS dataset ($Re_F=89$ for $M=256$ and $Re_F=358$ for $M=512$). For comparison, we also show the only statistically stationary simulation performed with an elongated domain, corresponding to $N=0.01$ ($Fr_F=0.89$, $Re_b=56$).

It is worth pointing out the important differences between the range of values taken up by $Fr_t$ and $Re_b$ in our DNS dataset. While $Re_b$ spans 1.5 orders of magnitude, ranging from $Re_b=32$ to $Re_b = 864$, the values of the turbulent Froude number are much more concentrated around $Fr_t \sim 1$. Specifically, the lowest value of $Fr_t$ is 0.44 while the highest is 1.27, both of which remain of order unity. One of the implications of this is that we have been unable to reach low values of $Fr_t$, $Fr_t \ll 1$, and so to access the strongly stratified turbulence (SST) regime, which is characterized by low values of $Fr_t$ and, concurrently, high values of $Re_b$ ($Re_b \gg 1$). The SST regime appears to be one of the most relevant regimes of stratified turbulence for geophysical applications \citep{rileylindborg2008}. Indeed, when we pushed the stratification levels beyond a certain value, thus reducing $Fr_F$ below the values given in figure~\ref{fig:fr}, the turbulence could not reach a steady state in which power input is balanced by dissipation and instead became temporally intermittent with large oscillations of the energy over time, as the flow alternated between laminar and turbulent phases. It is unclear whether such an intermittent state emerges because the SST regime is inherently inaccessible by stratified shear flows, as suggested by some previous studies \citep{zhou2017,smith2021}, or because higher resolutions are needed for the simulations to respect $Fr_t \ll 1$ together with $Re_b \gg 1$. We return to this issue in the final discussion, \S\ref{sec:conclusions}. 

We finish this section by considering the boundary between the statistically stationary runs presented in Figure~\ref{fig:fr} and the intermittent runs. A run at $M=256$ with $Re_F =89$ and $Fr_F = 0.28$ was performed, which showed temporal intermittency, alternating between laminar and turbulent states. 
As discussed in \S\ref{sec:sim}, the runs with $M=512$ also approach a regime transition when lowering $Fr_F$ below the values given in Figure~\ref{fig:fr}. Moreover, the breakdown of sustained turbulence into a temporally intermittent regime is also displayed by the runs in elongated domains, when stratification is increased too much. We will discuss the intermittent regime in more detail in \S\ref{subsec:intermittent}; for now we focus on the "last runs" at each resolution giving a steady state with sustained turbulence, i.e. the runs at lowest $Fr_F$ giving a steady state. Indeed, a pattern emerges when considering these runs, both in cubic and in elongated domains. In particular, such runs have values of $Fr_F$, $Fr_t$ and $Re_b$, respectively for $M=256$, $M=512$ and for the elongated domain, of $Fr_F = [0.34,\;0.16,\;0.89]$, $Fr_t = [0.62,\;0.44,\;0.29]$ and $Re_b = [32,\;47,\;56]$. It therefore appears that these runs have disparate values of $Fr_F$ but values of $Re_b$ and $Fr_t$ which are more close together. Particularly, the $Re_b$ curves shown in figure~\ref{fig:fr} appear to present their last data points just above a threshold around $Re_b \approx 30$. We therefore propose the existence of a critical value of $Re_b \approx 30$, at which there is a regime transition between sustained turbulence and an intermittent regime. This is of course a tentative regime boundary and it may indeed be possible that a similar boundary could be expressed in terms of $Fr_t$ (even though the evolution of $Fr_t$ and $Re_b$ as a function of $Fr_F$ given in figure~\ref{fig:fr} seems to be more consistent with a threshold based on $Re_b$). For the moment, suffice it to say that a regime boundary between sustained turbulence and intermittent regime described by a critical value of $Re_b$ is consistent with previous work on stratified shear flows. In the stratified inclined duct experiments of \citet{lefauve2019} a similar transition between an intermittent and a sustained turbulence regime was observed. As discussed by \citet{lefauve2019}, for high enough tilt angles, this regime transition was controlled by a parameter given by the Reynolds number times the tilt angle. \citet{lefauve2019} also found that at high tilt angles this parameter becomes proportional to a non-dimensional kinetic energy dissipation rate and, moreover, that it should asymptotically become proportional to $Re_b$. Therefore, results from a different configuration of stratified shear flow seem to be consistent with a critical value of $Re_b$ describing this regime transition.

\subsection{Interface thickness}
\label{subsec:interface}

\begin{figure}
\centering
\includegraphics[width=0.99\textwidth]{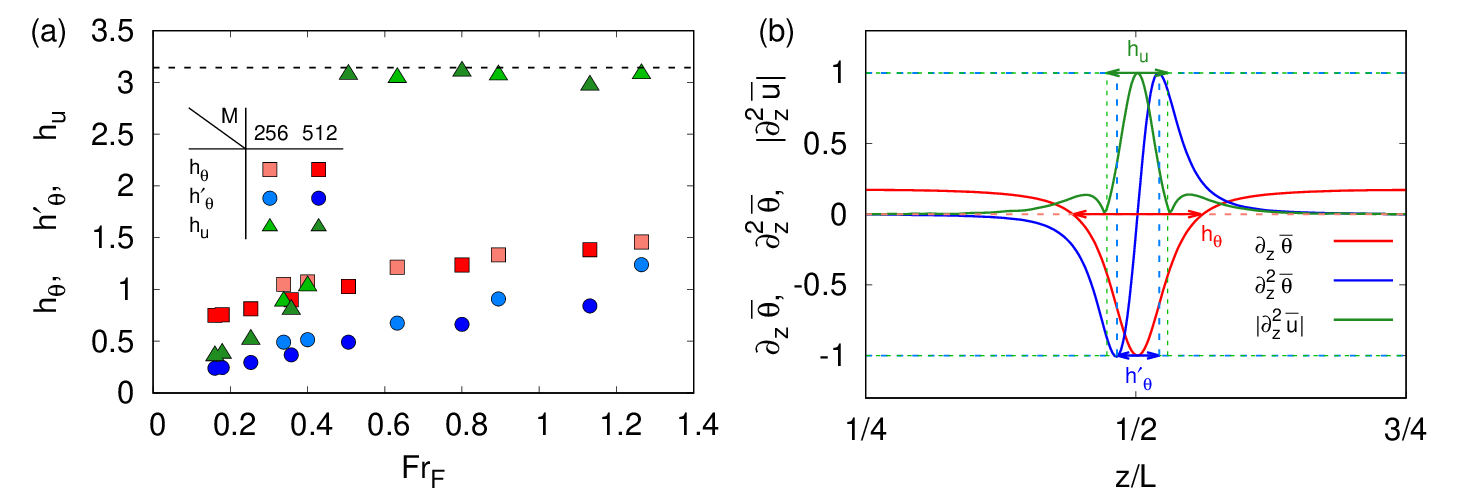}
\caption{Left panel: Interface thicknesses $h_\theta$, $h'_\theta$, and $h_u$ as functions of the forcing Froude number. Right panel: Visual example of the interface identification based on the vertical derivatives of the mean profiles $\overline{\theta}$ and $\overline{u}$ for the case $M = 512$, $N^2 = 4$.}
\label{fig:htheta}
\end{figure}

The pronounced layered structure observed in the mean density field, as visually described in Fig.~\ref{fig:profs} and discussed in \S~\ref{subsec:meanflow} and \S~\ref{subsec:rig}, motivates a detailed examination of the characteristic length scales that govern the interfaces. These scales are key for quantifying mixing across stable density interfaces and for understanding the spatial organization of stratified shear turbulence \citep[][]{turner1979}.

In stratified flows, the thickness of density interfaces has been long associated with diffusive processes. Early laboratory studies on density interfaces separating two turbulent layers of different density \citep[][]{crapper1974}, report two behaviours for the interface thickness: a diffusive behaviour at low P\'eclet number $Pe = u'\ell/\kappa$, where $u'$, $\ell$ are the velocity and length scales of the turbulence created in the layers, and a "turbulent" behaviour at high $Pe$. In the turbulent regime, the normalized interface thickness $h/\ell$ was an order unity constant, independent of $Pe$, while in the diffusive regime it was a decreasing function of $Pe$. A subsequent study by \citet{fernando1989} found that the data of \citet{crapper1974} were consistent with $h/\ell \sim Pe^{-1/2}$, meaning that the interface thickness obeys a diffusion-entrainment balance giving $h \sim \sqrt{\kappa \ell/u'}$, in which the molecular diffusion inside the interface is balanced by turbulent entrainment on either side of the interface. The stability analysis of stratified Kolmogorov flow carried out by \citet[][]{balmforth2002}, showed the development of internal boundary layers in the temperature field of the critical modes, whose height was found to scale approximately as $h \sim Pe^{-1/3}$ for highly viscous solutions with fixed $Re=1.92$, $Ri_g =0.01$ \citep[we refer the reader to][for the exact definitions of these parameters]{balmforth2002}. In fully turbulent conditions, however, it is expected that the interface thickness will scale differently, as for the turbulent regime of \citet{crapper1974}. The numerical study of \citet{smyth2000} considered the time evolution of a stratified shear layer with initially coincident hyperbolic tangent velocity and density profiles, which undergoes shear instability and transition to a fully turbulent state. They monitored the evolution of both shear layer thickness and density interface thickness and found that they scaled well with a lengthscale constructed from the total velocity and density change across the mixing layer. \citet{smyth2000} then put this lengthscale in relation to the Ozmidov scale and to other physical scales of their simulation. In the simulations of \citet{smyth2000}, the interface thickness therefore appears to not be controlled by diffusive processes but rather by a balance between inertia and buoyancy.

We probe the interface thickness behaviour in the present simulations. We evaluate several diagnostic measures, based on both the mean density profile $\overline{\theta}(z)$ and the mean velocity profile $\overline{u}(z)$, to quantify the interface geometry and its relation to the mean shear structure. The density interface is characterized by a strong gradient, appearing as pronounced peak in the profile of $\partial_z\overline{\theta}$. A natural geometric definition of its thickness, denoted $h_\theta$, is given by the distance between consecutive zeros of $\partial_z\overline{\theta}$, which mark the edges of the interface. The core of the interface corresponds to the peak of $\partial_z\overline{\theta}$. An alternative measure, $h'_\theta$, is based on the maxima of the second derivative $|\partial_{z}^2\overline{\theta}|$, which identify points of maximal curvature and thus provide an estimate of the inner interface width. 

The stationary solutions of our DNS, display a configuration in which the density interface separates two adjacent and oppositely signed shear layers. The centre of the density interface corresponds to a region of zero mean shear, in which the streamwise velocity $\overline{u}$ is maximum. As shown in Figure~\ref{fig:profs}, at high $N$ the velocity profile $\overline{u}(z)$ becomes "peaky", with the appearance of two symmetric inflection points. We can therefore form a velocity-based lengthscale $h_u$ by measuring the distance between these two inflection points, where $\partial^2_{z} \overline{u} = 0$. This quantity represents the spacing that forms in between the adjacent shear layers, i.e. it represents a low shear region sandwiched in between the shear layers and we therefore call $h_u$ the inter-shear-layer thickness. As discussed in \S\ref{subsec:meanflow} and \S\ref{subsec:rig}, we believe that the density structure during steady state results directly from the structure of the velocity field, particularly we have argued that the density interface should correspond to the regions of zero mean shear. If this picture is consistent, the interface thickness, quantified by $h_\theta$, $h'_\theta$, should be the same as the inter-shear-layer-thickness $h_u$. So the results presented herein serve also as a test on our physical understanding of the problem.

The comparison of these definitions, shown in Fig.~\ref{fig:htheta} as a function of the forcing Froude number $Fr_F$, reveals that although all measures remain of comparable order of magnitude, they display distinct trends and no universal scaling. For the density field, the different definitions yield consistent but quantitatively distinct results. The zero-crossing measure $h_\theta$ gives slightly larger values than the curvature-based estimate $h'_\theta$, indicating that it encompasses a broader geometric envelope of the interface. The velocity-based lengthscale is quantitatively similar to the density-based lengthscales at low $Fr_F$, giving $h_u \sim h_\theta$ and $h_u \sim h_\theta'$. This is consistent with our expectation that density interface thickness and inter-shear-layer-thickness should be controlled by the same dynamics. At higher $Fr_F$, however, $h_u$ ceases to have the same behaviour of $h_\theta$ and $h_\theta'$ and becomes constant, $h_u \approx \pi = L/2$. This is because, as can be seen from Fig.~\ref{fig:profs}, at weaker stratification the velocity profile becomes approximately piecewise linear, presenting inflection points only at the centre of each shear layer, and so $h_u$ latches on to these points, which are spaced by $L/2$, following the forcing structure. At high $Fr_F$, $h_u$ therefore stops being a representative measure of the thickness of the low-shear zone between the shear layers. It is therefore not surprising that $h_u \sim h_\theta$ and $h_u \sim h'_\theta$ break down.

As a final point, note that all three heights  remain of the same order as the buoyancy and Ozmidov lengthscales (see Table~\ref{tab:param}), consistent with a regime $Fr_t \simeq 1$, where no large scale separation between them is expected. Unlike diffusive or weakly turbulent staircases, the present configuration does not obey a diffusive scaling $h_\text{diff} \sim \sqrt{\kappa \tau}$. Indeed, it was attempted to scale $h_\theta$, $h'_\theta$ and $h_u$ with $h_\text{diff}$, using a timescale $\tau = N^{-1}$ and $\tau = u_{\rm rms}^{\prime 2}/\epsilon$, but this did not work.

\subsection{Vertical mass flux and turbulent mixing}
\label{subsec:flux}

Mixing in stably stratified turbulence plays a crucial role in controlling the vertical transport of momentum and buoyancy in the ocean interior. It is now well established that a large fraction of this mixing results from the dissipation of internal waves into small-scale turbulence \citep{gregg2018,delavergne2020}. Laboratory and numerical studies of stratified turbulence have shown that the efficiency of turbulent mixing depends strongly on the flow regime \citep{shih2005,maffioli2016,gregg2018}. The present simulations with periodic boundary conditions are able to capture a net upward mass flux through the numerical domain. Under statistically stationary conditions this mass flux is necessarily constant for every vertical level, as we will now see. We begin this section by focusing on this vertical mass flux and we later consider the mixing efficiency. 

\begin{figure}
\centering
\includegraphics[width=0.99\textwidth]{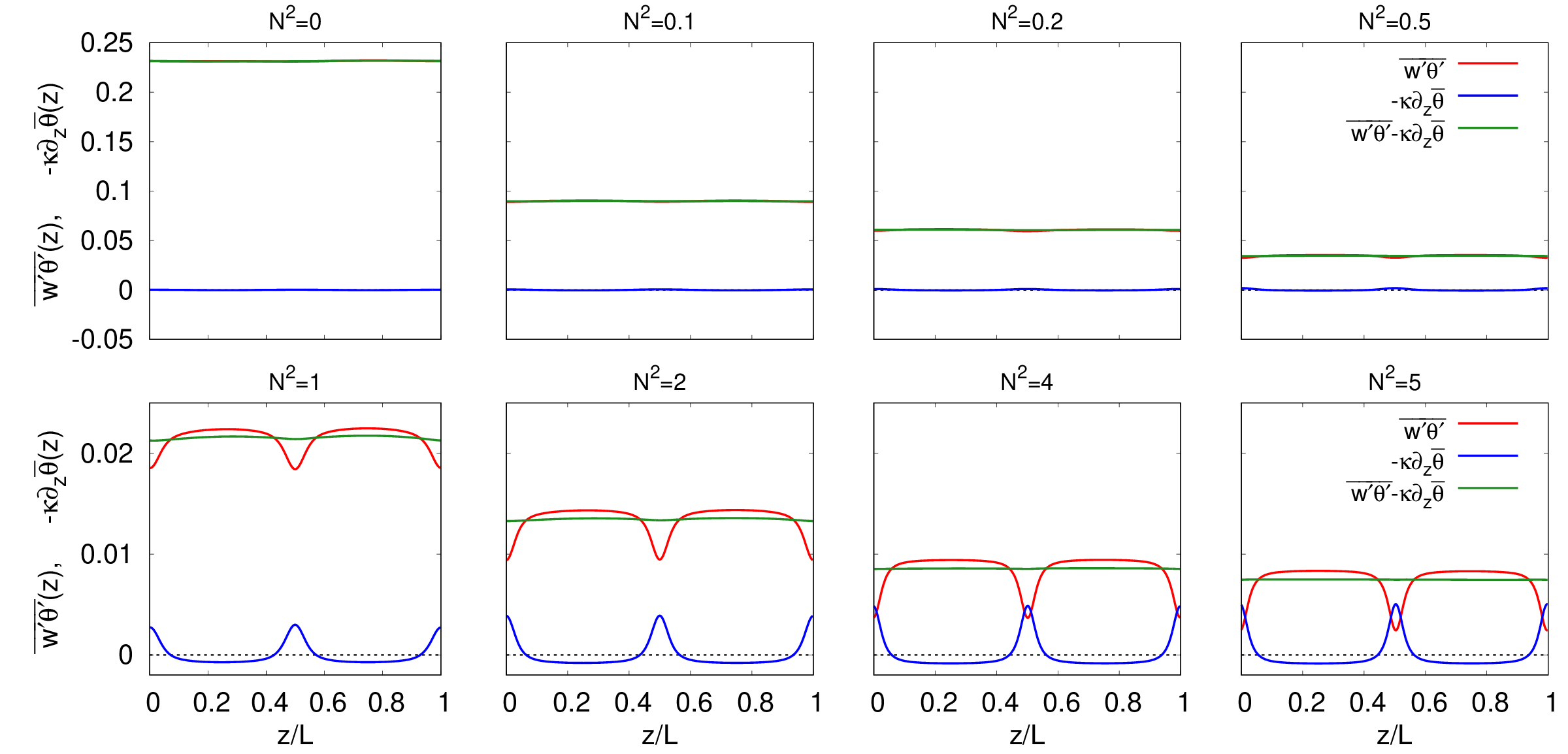}
\caption{Vertical profiles of the turbulent vertical mass flux $\overline{w'\theta'}$ (red) and the diffusive flux $-\kappa \, d_z \overline{\theta}$ (blue), along with the net upward flux given by their sum, for simulations at resolution $M=512$ with varying buoyancy frequency $N$.}
\label{fig:bflux}
\end{figure}

To obtain an expression for the average vertical mass flux in our simulations, we turn to the equation for $\overline{\theta}$, equation~\eqref{M4}, and integrate it vertically, assuming stationarity, i.e. setting $\partial_t \overline{\theta}=0$. Under this assumption, all mean variables depend only on $z$ and so $\partial_z \to d_z$. Vertical integration of $\eqref{M4}$ then gives

\begin{equation}
\overline{w'\theta'} - \kappa d_z \overline{\theta} = C, \label{massflux}
\end{equation}

\noindent where $C$ is a constant. The constant $C$ represents the mean upward mass flux through any horizontal plane of the simulation. If it were not constant, there would be mass accumulation or destruction in some slice of fluid of finite height, spanning through the domain in the horizontal directions, which is impossible if the simulation is statistically stationary. 

The important thing about equation~\eqref{massflux} is that it tells us that the vertical mass flux has two separate contributions, one due to turbulent transport, $\overline{w'\theta'}$, and the other due to diffusive transport, $-\kappa d_z \overline{\theta}$. The diffusive term is non-zero since $\overline{\theta}(z)\neq 0$, i.e. there is a mean modification of the background linear density profile, as seen in \S\ref{subsec:meanflow}. Vertical profiles of turbulent and diffusive transport, and of their sum, are given in figure~\ref{fig:bflux} for the runs with $M=512$. Based on their results, the simulations can be divided into two separate groups. For low stratification, $N^2 \leq 0.5$, the diffusive transport is virtually zero and the vertical mass transport is due entirely to turbulent transport. On the other hand, for high stratification, $N^2 \geq 1$, the picture becomes more complicated: positive diffusive transport starts occurring at $z/L = (0.5,1)$, where the density interfaces are, which is balanced by a small and negative diffusive transport over the rest of the domain (indeed $\int_0^L d_z \overline{\theta} \ dz = \overline{\theta}(L) - \overline{\theta}(0) = 0$ so that the diffusive profile must integrate to zero). 

A quite remarkable feature of the mass fluxes presented in Figure~\ref{fig:bflux}, is that for strong stratification ($N^2=4,5$) the diffusive transport overtakes the turbulent transport, at least in the density interfaces. This is a striking result for a turbulent flow, in which diffusive transport should be small and turbulent transport should dominate. What it implies is that the flow within the density interfaces is relaminarising, while the rest of the domain remains actively turbulent. This is a feature we already pointed out when looking at the flow visualizations of Figure~\ref{fig:sect_te_film}, where at $N^2=4$ the local dissipation highlights a segregation of the turbulence to the two "well-mixed" shear layers, while the flow in the strongly stratified density interfaces appears laminar. Based on this result, we attempted to scale the height of the density interface using a diffusion scale, $h_\theta \sim \sqrt{\kappa \tau}$. As discussed in \S\ref{subsec:interface}, these attempts were inconclusive and the interface thickness does not appear to scale as a diffusive scale for the DNS simulations even at high stratification.

We now turn to the energetics of the mixing process. We here present the vertical energy budgets for the total kinetic energy $(1/2)\overline{|\bm{u}^2|} = (1/2)|\overline{\bm{u}}|^2+(1/2)\overline{|\bm{u}'|^2}$ and for the total potential energy $(1/2)N^2 \overline{\theta^2} = (1/2) N^2 \overline{\theta}^2 + (1/2) N^2 \overline{\theta'^2}$. The energy budgets for the mean flow kinetic and potential energy and for the turbulent kinetic and potential energy following from the Reynolds decomposition introduced in \S\ref{subsec:rey} are  presented in the appendix \ref{app:a}. From the governing equations \eqref{eq1}-\eqref{eq2}, assuming stationarity, the horizontally averaged kinetic and potential energy budgets can be written as
\begin{eqnarray}
&& T_u(z) = P_{in}(z) - \varepsilon(z) - B(z), \label{energy1} \\[0.2cm]
&& T_\theta(z) = B(z) - \varepsilon_p(z), \label{energy2}
\end{eqnarray}
where $P_{in}(z) = \overline{\bm{u} \cdot \bm{f}} = \overline{u}(z)F\cos(Kz)$ is the energy input, $\varepsilon(z) = \nu \overline{|\bm{\nabla}\bm{u}|^2}$ and $\varepsilon_p(z) = N^2 \kappa \overline{|\nabla \theta|^2}$ are the kinetic and potential energy dissipation rates, and $B(z)=N^2 \overline{w\theta} = N^2 \overline{w'\theta'}$ is the buoyancy flux, using the fact that $\overline{w}=0$. The terms $T_u(z) =  \partial_z [ \overline{w(|\bm{u}|^2/2+p)} - \nu\partial_z\overline{|\bm{u}|^2/2} ]$ and $T_\theta(z) = N^2 \partial_z [ \overline{w\theta^2/2} - \kappa\partial_z\overline{\theta^2/2} ]$ represent the divergence of vertical energy fluxes, including turbulent transport due to Reynolds stresses, pressure transport and molecular viscous/diffusive transport, for kinetic and potential energy, respectively. Summing the two equations yields the total energy budget $T(z) = \varepsilon_T(z) + P_{in}(z)$ with $\varepsilon_T(z) = \varepsilon(z) + \varepsilon_p(z)$ and $T(z) = T_u(z) + T_\theta(z)$. For energy conservation, $\langle T(z) \rangle_z = 0$ and the power input balances the total dissipation $P_{in} = \epsilon + \epsilon_p$.

\begin{figure}
\centering
\includegraphics[width=0.99\textwidth]{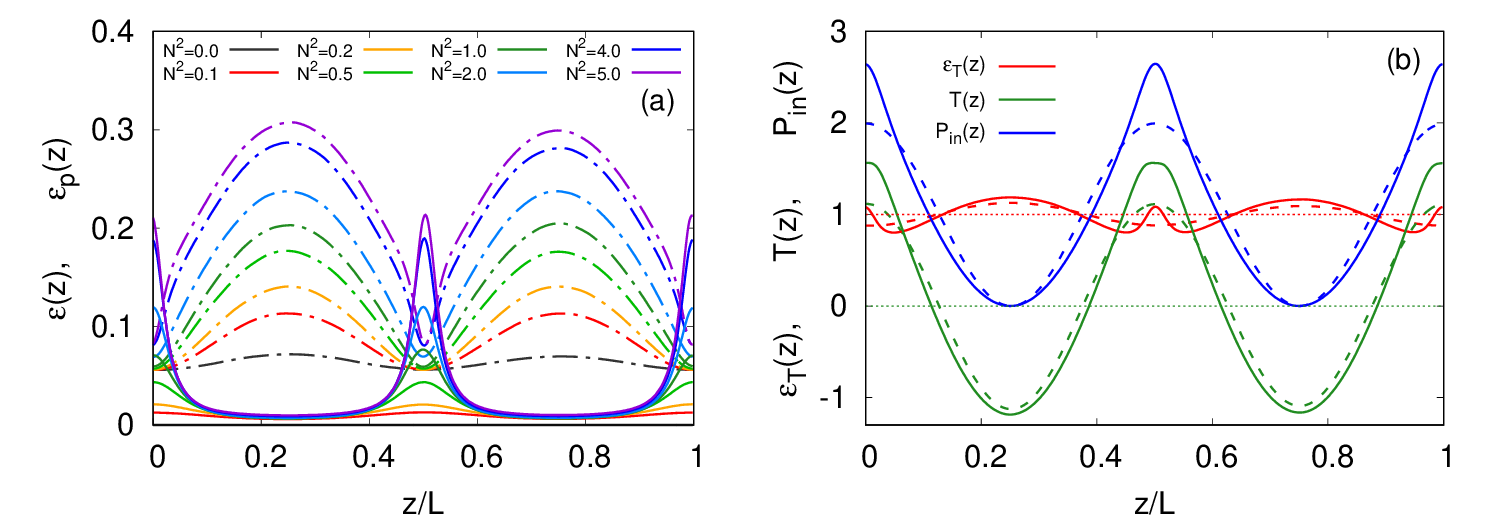}
\caption{Left panel (a): vertical profiles of kinetic energy dissipation rate $\varepsilon(z)$ (dash-dotted curves) and potential energy dissipation rate $\varepsilon_p(z)$ (solid curves) at varying stratification $N$ for resolution $M=512$. Right panel (b): profiles of total energy budget decomposed into total dissipation rate $\varepsilon_T(z)$, energy input $P_{in}(z)$ and energy transport fluxes $T(z)$ for case $N^2=0$ (dashed lines) and $N^2=4$ (solid lines) at $M=512$. All contributions are normalized by the energy input rate $P_{in}$.}
\label{fig:eps}
\end{figure}

In Figure~\ref{fig:eps} the different terms present in equation~\eqref{energy1}-\eqref{energy2} are represented as a function of $z/L$. The picture that emerges is that of an inhomogeneous flow. Figure~\ref{fig:eps}(a) shows the kinetic and potential energy dissipation rates $\varepsilon(z)$ and $\varepsilon_p(z)$. The kinetic energy dissipation $\varepsilon(z)$ is high in the well-mixed shear layers, whereas it falls dramatically in the density interfaces at $z/L = (0.5,1)$. This effect is more and more pronounced as the stratification is increased and is in agreement with the visualizations of $\varepsilon(x,z)$ in Figure~\ref{fig:sect_te_film}. Conversely, the potential energy dissipation $\varepsilon_p(z)$ has a strong peak in the density interfaces and drops to much lower values in the well-mixed shear layers. The interpretation is that the behaviour of $\varepsilon(z)$ and $\varepsilon_p(z)$ are linked by the large-scale structure of the flow. We have already discussed the fact that the turbulence and hence $\varepsilon(z)$ is active in the shear layers where $Ri_g \lesssim 1$ and the shear production can then efficiently generate turbulence. On the other hand, in the density interfaces $Ri_g \gg 1$ so that the flow is strongly stable, leading to relaminarisation with low levels of dissipation $\varepsilon(z)$. At the same time, the density stratification in the shear layers becomes significantly lower than $N^2$ in the shear layers as a result of turbulent mixing (this is why we call them "well-mixed" layers, even though some stratification survives as can be seen in Figure~\ref{fig:profs}), while it becomes much stronger than $N^2$ in the density interfaces. This results in smaller density fluctuations and smaller local density gradients, $| \nabla \theta| $, in the layers and much stronger density gradients in the interfaces, ultimately leading to $\varepsilon_p(z)$ being concentrated in the interfaces. 

In Figure~\ref{fig:eps}(b), the three terms present in the budget of total energy (kinetic + potential energy), obtained by summing equations~\eqref{energy1} and \eqref{energy2}, are represented. Again, an inhomogeneous picture emerges. The power input has a mode-2 shape, as a result of multiplying the mode-1 forcing term $F\cos(Kz$) with the mode-1 mean flow $\overline{u}(z)$. The transport term has a similar shape, except that it is translated to lower and negative values, which ensure that it integrates to zero. The positive values of $T(z)$ at $z/L = (0.5,1)$, where $P_\text{in}(z)$ is maximum, and the negative values at $z/L = (0.25,0.75)$, mean that on average $T(z)$ acts to transport total energy from $z/L = (0.5,1)$ towards $z/L = (0.25,0.75)$. The inhomogeneity is less marked for the total dissipation $\varepsilon_T(z)$, which has small excursions around a mean value of 1 (corresponding $\langle \varepsilon_T(z) \rangle_z =P_\text{in}$).

Moving to the global energetics, in stably stratified turbulence the exchange between kinetic and potential energy is mediated by the buoyancy flux $B$, which, in a statistically steady state and averaging over the entire fluid volume, equals the dissipation rate of potential energy, $\epsilon_p$ \citep{salehipour2015,sozza2015,gallon2024}. In the present setup $B = N^2 \langle w \theta \rangle = N^2 \langle w' \theta' \rangle$.
Together with the kinetic energy dissipation rate, $\epsilon$, these quantities provide a direct measure of the irreversible pathways through which the turbulent energy is removed. From these rates, we define the Nusselt number $Nu=1+\langle w' \theta' \rangle/\kappa$ \citep{grossmann2000}, quantifying the enhancement of scalar transport relative to molecular diffusion: $Nu=1$ corresponds to purely diffusive transport, while values $Nu > 1$ indicate increased mixing due to turbulent motions. Note that in the present setup the background gradient of $\theta$ is equal to $-1$ so that $\langle w'\theta' \rangle$ is equivalent to the turbulent diffusivity $\kappa_t$, i.e. $\kappa_t = \langle w'\rho'\rangle/\gamma = \langle w'\theta' \rangle$ (where $\gamma$ is the background density gradient). This means that the Nusselt number can also be written as $Nu = 1+ \kappa_t/\kappa$, which is a quantity that is often considered in the literature \citep[see, for example,][]{shih2005}. The turbulent diffusivity $\kappa_t$ is sometimes called the (turbulent) diapycnal diffusivity \citep{salehipour2015}. 

We consider additional dimensionless mixing parameters: the mixing coefficient $\Gamma = \epsilon_p / \epsilon$, measuring the relative importance of energy going into irreversible mixing compared to the energy being dissipated as kinetic energy, and the mixing efficiency $\eta = \epsilon_p/(\epsilon+\epsilon_p)$ representing the fraction of the total energy dissipation that is made up by irreversible mixing. Exploiting the relation $\langle w'\theta' \rangle = \epsilon_p /N^2$, one obtains an additional expression for $Nu$ as $Nu = 1 + \Gamma\, Pr\, Re_b$, where, in the present work, $Pr=1$. A similar expression was used by \citet{salehipour2015} for the turbulent diffusivity $\kappa_t$.

\begin{figure}
\centering
\includegraphics[width=0.75\textwidth]{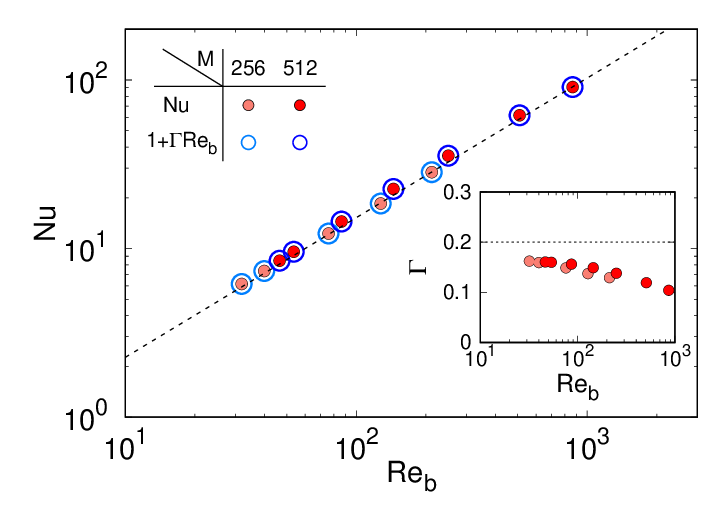}
\caption{Nusselt number $Nu=1 + \langle w'\theta' \rangle / \kappa$ (filled circles in red tones) as a function of the buoyancy Reynolds number $Re_b = \epsilon/(\nu N^2)$, compared with the relation $Nu = 1 + \Gamma Re_b$ (empty circles in blue tones). The fitted curve $Nu = \alpha Re_b^\beta$, with $\alpha = 0.4$ and $\beta = 0.8$, is indicated by a dashed line. In the inset, the mixing coefficient $\Gamma = \epsilon_p/\epsilon$ is plotted versus $Re_b$, with the Osborn–Cox benchmark value $\Gamma = 0.2$ for oceanographic applications shown as a dashed black line.}
\label{fig:nusselt}
\end{figure}

In Figure~\ref{fig:nusselt}, we plot $Nu$ as a function of the buoyancy Reynolds number $Re_b$. The data exhibit a clear scaling behaviour over a broad range of $Re_b$, approximately following a power-law $Nu \sim Re_b^\beta$ with exponent $\beta \simeq 0.8$. Figure~\ref{fig:nusselt} shows that as $N^2$ is increased and $Re_b$ is correspondingly decreased, the Nusselt number is decreased from relatively high values ($Nu \approx 90$) to values as low as $Nu \approx 6$. So as the stratification is increased, the ability of the turbulent flow to increase mixing beyond the laminar case is significantly reduced, which is consistent with increased flow stability. Note that this is a "non-dimensional statement" and it does not mean that the dimensional mixing of the density field reduces as stratification is increased. Indeed while $\langle w'\theta'\rangle $ reduces, the actual mixing, quantified by $\epsilon_p = N^2 \langle w'\theta'\rangle $, increases with increasing $N^2$. This is consistent with the total input power increasing linearly with $N$ (see \S\ref{subsec:peak}) and with the fact that at higher $N^2$ the background density differences are higher, increasing the potential to mix the density field. On the other hand, the present results for $Nu$ are consistent with the intuitive notion that increasing the stratification and the flow stability will reduce the mixing of a hypothetical passive scalar being added to the flow.  

Contrary to the results of other stratified turbulence configurations \citep{shih2005,maffioli2016}, the present results show that $\Gamma$ exhibits only a weak dependence on the flow parameters. When representing $\Gamma$ as a function of $Re_b$, as in the inset of Figure \ref{fig:nusselt}, small variations of $\Gamma$ are observed over 1.5 orders of magnitude in $Re_b$, leading to a correspondingly mild deviation of $Nu$ from a strictly linear dependence on $Re_b$. Indeed, the mixing coefficient $\Gamma$, and as a consequence also the mixing efficiency $\eta=\Gamma/(1+\Gamma)$, vary weakly throughout the entire set of simulations, with values close to the mean values $\Gamma \approx 0.12$ and $\eta \approx 0.1$, and approaching an asymptote for low values of $Re_b$ .  For reference, in oceanographic studies of fully developed stratified turbulence (i.e. $Re_b\gg 1$ and far from boundaries), an empirical benchmark first introduced by \citet{osborn1980} is to set $\Gamma$ around $0.2$ \citep{thorpe2007}. Our values are somewhat in the vicinity of this commonly assumed limit. High resolution DNS of sheared stratified turbulence with homogeneous shear were conducted by \citet{portwood2019}, who similarly found values of $\Gamma$ displaying a very weak dependence on $Re_b$, with an initial slight decrease followed by a plateau at around $\Gamma\approx 0.17$ for $Re_b > 200$.

\subsection{Transition to intermittent regime}
\label{subsec:intermittent}

Although the focus of this work is on the statistically stationary turbulent regime, we report here an observation of a transition toward an intermittent, oscillatory state occurring at long times in simulations with strong stratification. We believe this is the same intermittent regime found by \citet{garaud2016} in simulations of stratified Kolmogorov flow at low Prandtl number. To explore this regime transition occurring at strong stratification, i.e. low Froude number, we consider both the cubic domains and an elongated computational domain with resolution $M_x = 1024$, $M_y = M_z = 256$. This configuration retains the same large-scale forcing $F \cos(Kz)$ with $K=1$, but the increased horizontal extent allows a broader range of streamwise wavenumbers $k_x$ to develop, thus enhancing scale separation and accelerating the overall dynamics. Changing the aspect ratio of the domain can alter the numerical value of the critical Brunt–V\"ais\"al\"a frequency at which the transition to the intermittent regime occurs, but it does not change the qualitative nature of the transition itself. Rather, it primarily affects the timescales of the underlying dynamics.

\begin{figure}
\centering
\includegraphics[width=0.99\textwidth]{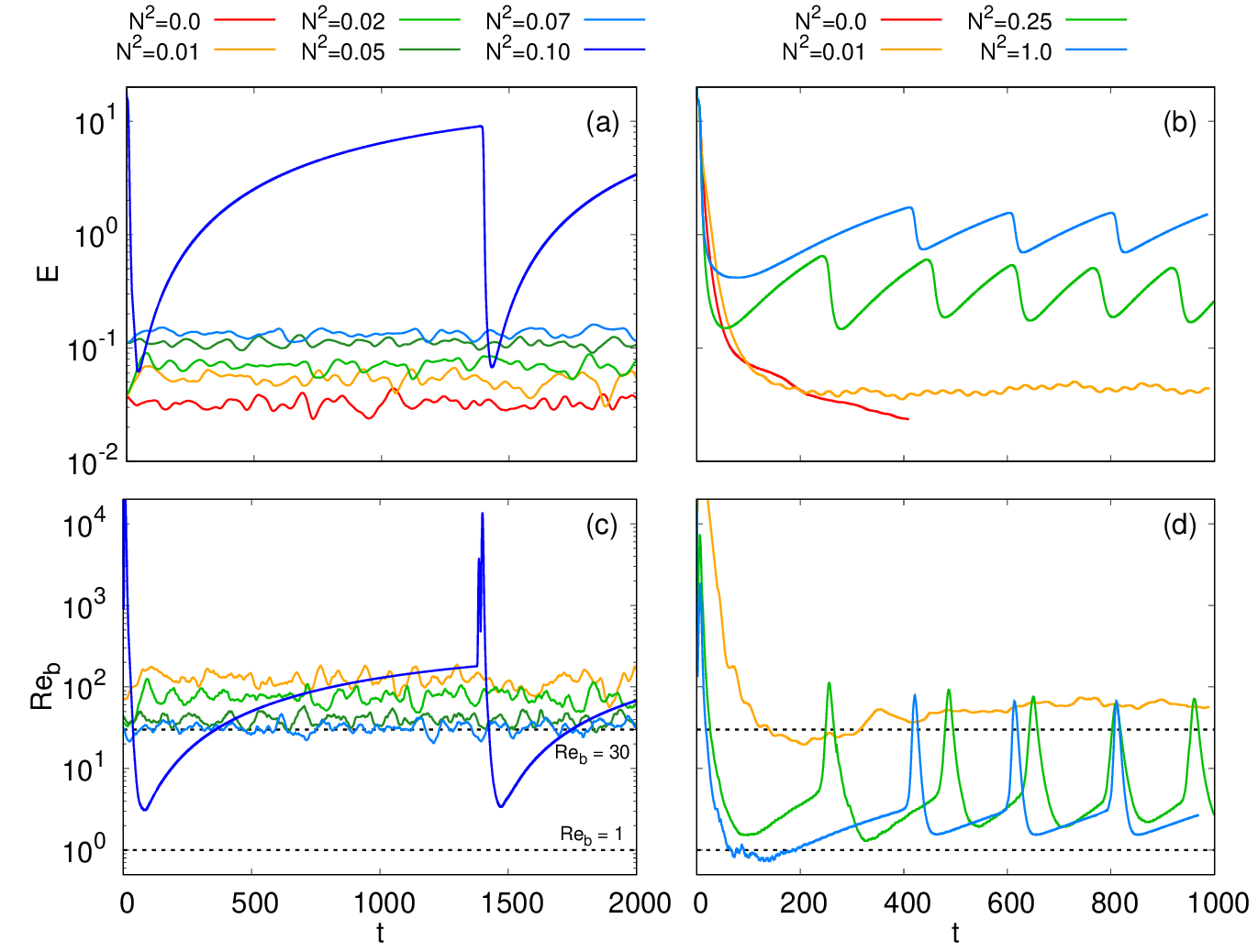} 
\caption{Time evolution of the total energy $E$ (panels a,b) and buoyancy Reynolds number $Re_b$ (panels c,d) for two domain geometries: cubic ($M=256$, panels a,c) and elongated ($M_x=1024$, $M_y=M_z=256$, panels b,d), and for different stratification. In both domains, the flow reaches a statistically steady state, with intermittent bursts and episodes of partial relaminarisation at high stratification. Dashed lines indicate critical values $Re_b=1$ and $Re_b=30$.}
\label{fig:ts}
\end{figure}

Figure~\ref{fig:ts}(a)-(b) shows the time evolution of the total energy $E$ for different values of the stratification parameter $N^2$. Panel~(a) reports simulations performed in a cubic domain at resolution $M = 256$. For weak stratification ($N^2 < 0.10$), the system reaches a statistically steady turbulent state whose mean energy level increases with increasing $N^2$. At $N^2 = 0.10$, however, the flow displays a distinctive intermittent behaviour characterized by a slow energy build-up followed by a rapid release, indicating a cyclic alternation between quasi-laminar and turbulent phases. 
Panel~(b) shows simulations performed in an elongated domain, where the transition to the intermittent regime becomes more evident, marked by recurrent bursts and partial relaminarisation. For the moderately stratified case ($N^2 = 0.01$), the energy begins to exhibit regular oscillations, suggestive of a crossover regime between sustained turbulence and laminarisation, in which internal waves start to play a significant role while a statistically steady state remains attainable.

\begin{figure}
\centering
\includegraphics[width=0.99\textwidth]{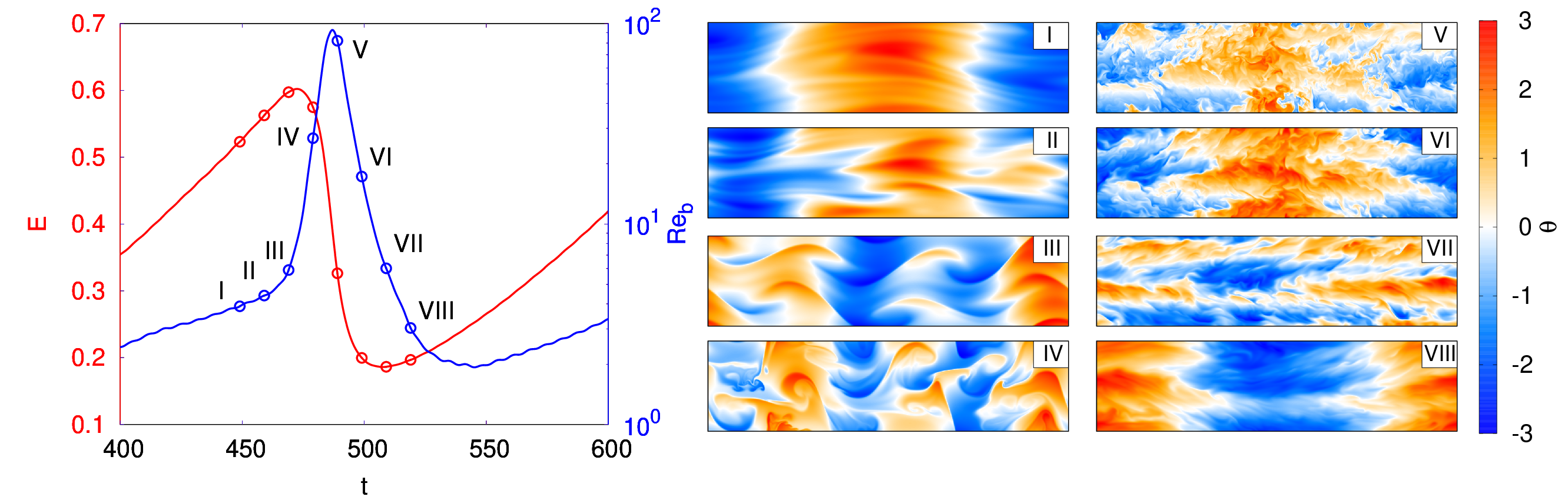}
\caption{Sequence of vertical sections of the density field $\theta(x,z)$ at different times (panels I--VIII) during a full oscillation cycle of the total energy $E$ (red curve) and buoyancy Reynolds number $Re_b$ (blue curve), whose time series are shown in the left panel. Empty circles indicate the instants corresponding to each snapshot. The sequence illustrates the abrupt transition from a laminar state (I--II), to the onset of Kelvin--Helmholtz-like instabilities (III--IV), followed by the development of turbulence (V--VI), and the subsequent decay into a relaminarised state with residual turbulent patches (VII--VIII). Parameters: $N^2 = 0.25$, $M_x=1024$, $M_y=M_z=256$.}
\label{fig:seq}
\end{figure}

Figure~\ref{fig:ts}(c)-(d) the evolution of $Re_b$ over time is shown. For both cubic and elongated domains, the stationary runs approach the value $Re_b=30$ when increasing $N^2$, which is our proposed regime boundary demarcating sustained turbulence from temporal intermittency. Interestingly, the temporal evolution of $Re_b$ in the intermittent runs shows that $Re_b$ crosses this boundary at different times in the simulations. As will be confirmed visually in figure~\ref{fig:seq}, where we zoom in on the flow behaviour during an individual cycle, the phases with high $Re_b$ ($Re_b > 30$) correspond to the turbulent bursts, while the low-$Re_b$ phases ($Re_b < 30$) correspond to the laminar phases, where $Re_b$ falls to values not far above $Re_b=1$. Comparing the behaviour of $Re_b(t)$ in the simulations in cubic and elongated domains, notable differences include the fact that $Re_b$ in the cubic domain reaches extremely high values during the turbulent burst ($Re_b \sim 10^4$), while in the elongated domain the maximum values are much lower, $Re_b \sim 100$.      

A closer inspection of a single oscillation cycle in the simulation in the elongated domain with $N^2 = 0.25$ is shown in Figure~\ref{fig:seq}. The left panel displays a zoom-in on the time series of $E(t)$ and $Re_b(t)$, with circles marking the times corresponding to the snapshots of the scalar field $\theta(x,z)$ shown in panels~I--VIII. At the beginning of the cycle (I--II), the flow is laminar and spatially organized. As time progresses (III--IV), the shear intensifies and triggers the onset of Kelvin--Helmholtz-like instabilities. These instabilities rapidly develop into a turbulent burst (V--VI), producing enhanced mixing and fine-scale structures. Subsequently, turbulence decays and the system relaminarises (VII--VIII), completing a full oscillation cycle characterized by a slow energy build-up followed by a fast release. In terms of $Re_b$, its maximum ($Re_b \sim 100$) corresponds to panel~V, which is the first panel after the breakdown into turbulence and appears visually to be the moment when the turbulence is most vigorous. Lower values of $Re_b$ ($Re_b < 10$) are associated with the initial laminar phases (I--III) and the final phases of relaminarisation (VII--VIII). Focusing on the zoom-in of $E(t)$ and $Re_b(t)$, it is also clear that the maximum of $Re_b$, which corresponds to the maximum of the dissipation $\epsilon$, corresponds to the rapid decrease of $E(t)$ in time, as the enhanced dissipation during the turbulent bursts leads to a rapid destruction of the total energy. 

The flow seems to exhibit a spatial modulation along the streamwise direction $x$, as illustrated in Figure~\ref{fig:seq} (see in particular panels I--II and VIII). This indicates a partial loss of translational invariance, a phenomenon that has also been reported in previous studies of Kolmogorov flows \citep[e.g.][]{sarris2007}, where increasing the domain aspect ratio was shown to promote the emergence of large-scale flow modulations. In the present simulations, this effect does not alter the qualitative behaviour of the flow nor the observed transition to the intermittent regime, but it highlights that the assumption of streamwise homogeneity may not hold strictly in elongated domains. The implications of this symmetry breaking for the long-term dynamics remains an open question.

The alternation between a slow build-up and a rapid release of kinetic energy suggests a self-regulating mechanism. In this regime, energy accumulates until the flow becomes locally unstable, triggering a turbulent burst that rapidly dissipates the stored energy. Similar cyclic or intermittent dynamics have been reported in previous numerical studies of sheared stratified turbulence, notably by \citet{garaud2016}, who investigated the regime at low Prandtl number for astrophysical flows, and by \citet{chung2012}, who observed sustained oscillations in homogeneous stratified shear turbulence at unity Prandtl number. Related behaviour has also been documented in laboratory experiments of stratified shear flows in inclined ducts \citep{lefauve2019,duran2023}, as well as in atmospheric observations of intermittent bursting within the stable boundary layer \citep{vanderlinden2020}. Our results, obtained at unity Prandtl number, thus complement these studies by revealing analogous mechanisms in a different configuration, contributing to a broader understanding of the dynamical phase space of sheared stratified turbulence. The oscillation period appears to increase with stratification strength $N^2$, although a quantitative study of this and other features of the temporally intermittent regime is left for future work.

\section{Discussion and conclusions}
\label{sec:conclusions}

We return here to the transition between the sustained turbulence regime and the temporally intermittent regime. Based on the present DNS results, we have proposed that this transition occurs at $Re_b \approx 30$. Following the analysis first presented by \cite{billant2001,brethouwer2007}, the buoyancy Reynolds number is a Reynolds number that describes the importance of viscous effects in terms of the vertical scale of the turbulent layers present in stratified turbulence. This Reynolds number is therefore expected to be particularly relevant for the SST regime, in which the flow is strongly anisotropic, but it has become customary in the literature to use it even under weak stratification, in cases with $Fr_t \gtrsim 1$ or even with $Fr_t \gg 1$ \citep{shih2005,salehipour2015,portwood2019}. A critical value of $Re_b = 30$ is therefore consistent with the transition from a fully turbulent regime (for $Re_b > 30$) to a regime that is affected by viscosity (for $Re_b < 30$), which occurs because $Re_b \gg 1$ is not respected below this value. Intuitively, the intermittent regime should be affected by viscosity, since its turbulent bursts die out quickly, presumably under the action of viscosity, leaving way to extended periods of laminar flow. In this sense, the current transition is not dissimilar to the transition occurring in homogeneous stratified turbulence between the SST regime and the viscosity-affected regime \citep{brethouwer2007}, albeit with the extra ingredient of temporal intermittency, which is not present in homogeneous stratified turbulence. As demonstrated by a detailed numerical study on the transition between these two regimes of homogeneous stratified turbulence \citep{bartello2013}, this transition occurs around $Re_b \approx 10$. Moreover, the study of a temporally evolving stratified shear layer of \citet{smyth2000} identified a value $Re_b = 20$, as the value below which viscosity started being important. It therefore appears that for different classes of stratified turbulent flows a transition towards a regime affected by viscosity occurs for $Re_b = O(10)$. In the present configuration, if $Re_b$ is pushed further down to $Re_b \lesssim 1$ further regime changes may occur, and we may arrive at a fully laminar flow. It is clear that more work is necessary, both on the passage from stationary turbulence to intermittent turbulence and on the intermittent regime itself, in order to fully elucidate the regime transitions taking place in stratified Kolmogorov flow. Finally, note that we have not explored a potential Prandtl number dependence of this regime boundary, which would be consistent with the important role that molecular diffusion of $\theta$ starts playing in the runs approaching regime transition (i.e. the most strongly stratified runs), as highlighted by the results of \S\ref{subsec:flux}. The fact that $Pr=1$ across our DNS dataset does not allow us to explore such a $Pr$-dependence. Further investigation is needed on this important point.

It is now well understood that homogeneous stratified turbulence can be in different regimes depending on the values of the turbulent Froude number, buoyancy Reynolds number and turbulence Reynolds number $Re_t$ \citep{billant2001,lindborg2006,brethouwer2007}. As already mentioned, a regime of particular interest for its potential geophysical applications is the SST regime, which requires $Fr_t \ll 1$ together with $Re_b \gg 1$  Now, the stratified turbulence regimes based on $Fr_t$, $Re_t$, $Re_b$ were obtained using theoretical considerations valid for homogeneous stratified turbulence without a mean flow. The present case is somewhat different as there is a mean flow that imposes a mean shear on the flow, which is also the production mechanism of the turbulence. This may shape the stratified turbulence that develops making turbulent stratified shear flows different from homogeneous stratified turbulence. This is what has been argued in recent literature, claiming that stratified shear flows cannot enter the SST regime \citep{zhou2017,smith2021}. In particular, the work of \citet{zhou2017} focused on stratified plane Couette flow in the turbulent regime and found that $Fr_t \sim 1/\sqrt{Ri_g}$ in their simulations. This implies that the turbulence of this stratified shear flow cannot be strongly stratified, $Fr_t \ll 1$, since for shear production $Ri_g \lesssim 1$ is required and hence $Fr_t \gtrsim 1$. 

In the present DNS we have been unable to reach the SST regime and have essentially remained in regimes described by $Fr_t\sim 1$, while moving between stationary and a temporally intermittent regimes based on the value of $Re_b$. We have therefore found no evidence that contradicts the claim that the SST regime is inaccessible for stratified shear flows. We wish however to suggest the general idea that vertical shear production is not \textit{per se} incompatible with strongly stratified conditions. Shear production requires $Ri_g \lesssim 1$, which puts a constraint on the vertical shear and so on the vertical lengthscale $\ell_v$ of the flow. However there is no constraint on the horizontal lengthscale of the flow, $\ell_h$, which controls the turbulent Froude number since $Fr_t \sim |\bm{u}'|_{\rm rms}/N\ell_h$ \citep[see, e.g.,][]{maffioli2016}. Moreover, in stratified Kolmogorov flow, we expect $\ell_v$ to be set by the wavenumber of the forcing, $\ell_v \sim 1/K \sim L_z$ for $K=1$, where $L_z$ is the height of the box. Turbulence in the SST regime is known to be highly anisotropic with $\ell_h \gg \ell_v$. This means that $\ell_h \gg L_z$ is required, which is not possible in cubic domains. Thus, in order for the SST regime to be accessible for the simulations, it is necessary to move to "rectangular" domains, with $L_x, L_y \gg L_z$. The present runs in elongated domains, with $L_x = 8\pi$, $L_y =2\pi$, $L_z=2\pi$, are a step in this direction. Note that in the stationary run in this elongated domain we achieved the lowest value of $Fr_t$ throughout the DNS dataset, $Fr_t =0.29$, though not sufficiently low to reach $Fr_t \ll 1$. More work investigating stratified shear flow in anisotropic domains is therefore needed in order to elucidate this issue. It may indeed be an interesting point to consider also for other configurations of stratified shear flow, such as wall-bounded flows or isolated mixing layers.

One of the main findings of the present work is that a marked layer-interface structure emerges naturally from the dynamics as the stratification is increased. Two layer-interface sequences are observed over the height of the box, as a result of the forcing wavenumber being $K=1$, with the density interfaces located at $z/L = 0.5,\,1$, where the mean shear $d_z \overline{u}$ vanishes. Given the initially constant stratification in our system, the density interface is free to choose its preferred location along the vertical. On the other hand, several previous studies on stratified shear layers use the classical setup of co-located shear and density interfaces \citep{smyth2000,salehipour2015,lefauve2019,smith2021}. The classical setup may be most relevant for two-layer exchange flows, as found at river confluences or in estuarine flows. In many geophysical scenarios, however, there is no constraint on shear and density gradients being co-located. This was emphasized by a recent study on asymmetric shear layers, in which shear and density gradient were offset, leading to different dynamics, even though the study reports a tendency for the shear and density gradient to become co-located in their DNS of the problem \citep{olsthoorn2023}. The present study starts with no initial density interface and shows a natural tendency for interfaces to form where the shear is minimum, while relatively well-mixed layers form where the shear is maximum. We offered a rationalization for this behaviour based on the susceptibility of the flow to shear instability.  
The spontaneous localization of density interfaces at points of vanishing shear has already been observed in wall-bounded stratified shear flows, in particular in channel flows in which a density interface emerges at channel mid-height where, by symmetry, the mean shear is zero \citep{zonta2012,cen2024}. It may be a more general feature for those stratified shear flows in which the density profile can evolve freely.

\section{Acknowledgments}
Both authors acknowledge PSMN (P\^ole Scientifique de Mod\'elisation Num\'erique, ENS de Lyon) for computing resources. A.S. acknowledges support from 2021-2023 post‑doctoral fellowship program LABEX MILYON (ANR‑10‑LABX‑0070) of Universit\'e de Lyon, within the Investissements d’Avenir program (ANR‑11‑IDEX‑0007) operated by the French National Research Agency (ANR). A.S. also acknowledges the support of Fondazione Compagnia di San Paolo under the framework of the TRAPEZIO Call – \emph{"Paving the way to research excellence and talent attraction"}, Line 2: MSCA Seal of Excellence – Second Edition (ROL ID: 124142). A.S. thanks G. Boffetta, S. Musacchio, F. De Lillo, M. Cencini and A.S. Lanotte for early suggestions and stimulating discussions.

\section{Declaration of Interest} 
The authors report no conflict of interest.

\appendix

\section{Equations for the turbulent fluctuations and energy budget}
\label{app:a}

We consider here the Reynolds decomposition introduced in section~(\ref{subsec:rey}). Equations for the fluctuations are obtained subtracting mean flow equations eqs.~\eqref{M1}--\eqref{M4} from the complete Boussinesq equations~\eqref{eq1}--\eqref{eq2}. We obtain

\begin{eqnarray}
\partial_t \bm{u}' + \overline{\bm{u}}\cdot\nabla \bm{u}'
+ w'\partial_z \overline{\bm{u}} + \bm{u}'\cdot\nabla \bm{u}' 
- \partial_z \overline{ w'\bm{u}' } &=& -\bm{\nabla} p' 
- N^2 \theta'\widehat{\bm{z}} + \nu \bm{\nabla}^2 \bm{u}', \\[0.2cm]
\partial_t \theta' + \overline{\bm{u}}\cdot\nabla \theta' 
+ w'\partial_z \overline{\theta} + \bm{u}'\cdot\nabla \theta' 
- \partial_z \overline{ w'\theta' } &=& w' + \kappa \bm{\nabla}^2 \theta'.
\end{eqnarray}
Incompressibility applies as well as for velocity fluctuations, i.e. $\bm{\nabla}\cdot\bm{u}'=0$.

We derive here the energy balance for the horizontally averaged mean flow and for the fluctuations. We multiply the mean momentum equation by $\overline{\bm{u}}$ and the mean scalar equation by $N^2 \overline{\theta}$, and average in the horizontal directions. This yields the mean energy equations:
\begin{eqnarray}
\partial_t \, \tfrac{1}{2}|\overline{\bm{u}}|^2
+ \partial_z (\overline{\bm{u}} \cdot \overline{w'\bm{u}'}) 
-  \overline{ w' \bm{u}' } \cdot \partial_z \overline{\bm{u}} &=& \tfrac{\nu}{2} \partial_{zz} |\overline{\bm{u}}|^2 - \nu |\partial_z \overline{\bm{u}}|^2 + \overline{\bm{u}}\cdot \bm{f}, \\[0.3cm]
\partial_t \, \tfrac{1}{2}N^2 \overline{\theta}^2
+ N^2 \partial_z (\overline{\theta} ~ \overline{w'\theta'})
- N^2 \overline{ w' \theta' } \partial_z\overline{\theta}
&=& \tfrac{\kappa}{2} N^2 \partial_{zz} \overline{\theta}^2 - \kappa N^2 |\partial_z \overline{\theta}|^2 .
\end{eqnarray}
We have used the vectorial identity for a generic solenoidal field $\bm{A}\cdot\nabla^2\bm{A} = \tfrac{1}{2}\nabla^2 |\bm{A}|^2 - |\bm{\nabla} \bm{A}|^2$. These relations describe the evolution of mean kinetic and potential energy. They include the following terms: $P_{\bm{u}} = - \overline{ {\bm{u}'w'} } \partial_z \overline{\bm{u}} $ and $ P_{\theta} = - \overline{ {w'\theta'} } \partial_z \overline{\theta}$ are the Reynolds-stress production terms and represent the transfer of momentum and density from the mean fields to the fluctuating fields, $T_{\overline{\bm{u}}} =  \partial_z [ \overline{\bm{u}}\cdot\overline{w'\bm{u}'} - \tfrac{\nu}{2} \partial_{z} |\overline{\bm{u}}|^2]$ and $T_{\overline{\theta}} = N^2 \partial_z [ \overline{\theta}\cdot\overline{w'\theta'}-\tfrac{\kappa}{2} \partial_{z} \overline{\theta}^2 ]$ are the viscous and diffusive transport fluxes, $\overline{\varepsilon} = \nu |\partial_z \overline{\bm{u}}|^2$ and $\overline{\varepsilon_p} = \kappa N^2 |\partial_z \overline{\theta}|^2$ are the dissipations by viscosity and diffusivity, and $P_{in} =\overline{\bm{u}}\cdot \bm{f}$ is the energy rate injected by the forcing. Notice that no buoyancy exchange term does appear, so there's no coupling between density and velocity in the mean energy equations, as mean vertical velocity is zero $\overline{w}=0$.

For the fluctuating fields, we proceed similarly. Multiplying the fluctuation momentum equation by $\bm{u}'$ and the scalar equation by $N^2\theta'$ and averaging, we obtain:
\begin{eqnarray}
\partial_t \tfrac{1}{2} \overline{ |\bm{u}'|^2 }
+ \overline{ \bm{u}'w' } \partial_z \overline{\bm{u}}
+ \partial_z \tfrac{1}{2} \overline{ {w'|\bm{u}'|^2} }
= -\partial_z \overline{ {w'p'} } 
- N^2 \overline{ {w'\theta'} } 
 + \tfrac{\nu}{2} \partial_{zz} \overline{ {|\bm{u}'|^2} }
- \nu \overline{ {|\nabla \bm{u}'|^2} }, \\[0.2cm]
\partial_t \tfrac{1}{2} N^2 \overline{ {\theta'^2} }
+ N^2 \overline{ {w'\theta'} } \partial_z \overline{\theta}
+ \tfrac{1}{2} N^2 \partial_z \overline{ {w'\theta'^2} }
= N^2 \overline{ {w'\theta'} } + \tfrac{\kappa}{2} N^2 \partial_{zz} \overline{ {\theta'^2} } 
- \kappa N^2 \overline{ {|\nabla \theta'|^2} }.
\end{eqnarray}

We define the following terms. The quantities $P_{\bm{u}} = -\,\overline{u'w'}\,\partial_z \overline{u}$ and $P_{\theta} = -\,\overline{w'\theta'}\,\partial_z \overline{\theta}$ are the shear- and buoyancy-production terms of turbulent kinetic and potential energy. The same terms appear in the mean-flow energy budget with opposite sign, acting as sinks of mean kinetic and potential energy while serving as source terms in the fluctuation budgets. The buoyancy flux $B = N^2\,\overline{w'\theta'}$ represents the reversible exchange between turbulent kinetic and potential energy. The terms $\tfrac{1}{2}\partial_z \overline{w'|\bm{u}'|^2}$ and $\tfrac{1}{2}\partial_z \overline{w'\theta'^2}$ are the turbulent transport fluxes, while $\partial_z \overline{w'p'}$ is the pressure-transport contribution. The terms $\tfrac{\nu}{2}\partial_z \overline{|\bm{u}'|^2}$ and $\tfrac{\kappa}{2}\partial_z \overline{\theta'^2}$ are the viscous and diffusive transport fluxes, and $\nu\,\overline{|\nabla \bm{u}'|^2}$ and $\kappa\,N^2\,\overline{|\nabla \theta'|^2}$ are the viscous and diffusive dissipation rates. 

The vertical energy transport fluxes can be reorganized as:
\begin{eqnarray}
T_{u'}(z) &=& \partial_z \left[ \tfrac{1}{2}  \overline{ w' |\bm{u}'|^2 } + \overline{ w' p' } 
- \tfrac{\nu}{2} \partial_z \overline{ {| \bm{u}' |^2} } \right], \\
T_{\theta'}(z) &=& N^2 \partial_z \left[ \tfrac{1}{2} \overline{ w' \theta'^2 } - \tfrac{\kappa}{2} \partial_z \overline{ \theta'^2 } \right].
\end{eqnarray}

The final set of the energy budget equations become
\begin{eqnarray}
\partial_t \, \tfrac{1}{2}|\overline{\bm{u}}|^2 
+ T_{\overline{\bm{u}}} 
&=& -P_{\bm{u}} -\overline{\varepsilon} + P_{in}, \label{eq:a9} \\[0.2cm]
\partial_t \, \tfrac{1}{2}N^2 \overline{\theta}^2 
+ T_{\overline{\theta}} 
&=& -P_{\theta} - \overline{\varepsilon}_p, \label{eq:a10} \\[0.2cm]
\partial_t \tfrac{1}{2} \overline{ |\bm{u}'|^2 } 
+ T_{u'} 
&=& P_{\bm{u}} - B - \varepsilon', \label{eq:a11} \\[0.2cm]
\partial_t \tfrac{1}{2} N^2 \overline{\theta'^2} 
+ T_{\theta'} 
&=& P_{\theta} + B - \varepsilon_p'. \label{eq:a12}
\end{eqnarray}

By summing the mean and fluctuation budgets, Eqs.~\eqref{eq:a9}--\eqref{eq:a12}, we obtain the horizontally averaged budget for the total energy profile, $E=\tfrac{1}{2}\overline{|\bm{u}|^2} + \tfrac{1}{2}N^2\overline{\theta^2}$:
\begin{eqnarray}
\partial_t E(z) + T(z) = P_{in}(z) - \varepsilon(z) - \varepsilon_p(z).
\end{eqnarray}
This equation shows that the total energy evolves due to external forcing, is redistributed vertically by transport, and is ultimately removed by viscous and diffusive dissipation. 

Note that when summing the mean and fluctuation budgets to form the total energy budget, the production terms cancel out. They represent internal exchanges between the mean flow and fluctuations and do not contribute to the evolution of the total energy. We also remark that, when the total energy budget is integrated over $z$, the transport fluxes contribute no net change in the total energy, because of the periodic boundary conditions (i.e. $\langle T(z) \rangle_z = 0$).

Finally, assuming the system reaches a statistically steady state, we can set $\partial_t E=0$, and obtain the exact expression used to compute the energy budget in Eqs.~\eqref{energy1}-\eqref{energy2} and shown in Figure~\ref{fig:eps}.

\section{Loss of monochromaticity}
\label{app:b}

Beyond the amplitude estimate based on the peak velocity, i.e. $U = (\overline{u}_{\max} - \overline{u}_{\min})/2$, one can adopt more refined statistical measures that capture the shape of the mean profile.

In the ideal Kolmogorov case the mean flow is sinusoidal, \(\overline{u}(z) = U \cos(Kz)\). This monochromatic profile contains a single Fourier mode, and its volume-averaged square is \(\langle \overline{u}(z)^2 \rangle = U^2/2\), since the average of \(\cos^2\) over one period is \(1/2\). Any deviation from this reference form indicates the presence of higher harmonics and thus a loss of monochromaticity.

This motivates the use of non-monochromaticity indices. For any periodic mean profile $\overline{\psi}(z)$ with amplitude $\Psi = (\max\overline{\psi} - \min\overline{\psi})/2$, we define the normalized profile $\tilde{\psi}(z) = \overline{\psi}(z)/\Psi$ and introduce the dimensionless statistical moments
$$
\mu_n[\tilde{\psi}] = \frac{1}{2\pi} \int_0^{2\pi} \tilde{\psi}(z)^n \, dz.
$$

The second moment $\mu_2[\tilde{\psi}]$ acts as a non-monochromaticity index, quantifying the deviation from a pure sinusoidal shape. For a cosine profile, $\mu_2 = 1/2$. Sharper or more distorted profiles yield $\mu_2 \approx 1/3$, indicating significant deviation from monochromaticity and a broader spectral content.

To further characterize the shape of the mean profiles, one may consider higher-order moments. The skewness and kurtosis of $\tilde{\psi}$ are defined as 
$$
\mathcal{S}[\tilde{\psi}] = \mu_3[\tilde{\psi}]/ \mu_2[\tilde{\psi}]^{3/2}, \qquad 
\mathcal{K}[\tilde{\psi}] = \mu_4[\tilde{\psi}]/ \mu_2[\tilde{\psi}]^2.
$$

The skewness and kurtosis provide complementary information on the shape of the normalized profiles. The skewness measures the imbalance between positive and negative excursions of the signal, while the kurtosis increases when the profile develops sharper gradients or more peaked structures. For reference, a pure cosine profile has zero skewness and kurtosis $\mathcal{K} = 3/2 = 1.5$, while a piecewise-linear sawtooth of the same amplitude has zero skewness and a larger kurtosis, $\mathcal{K} = 9/5 = 1.8$. These values therefore provide convenient limits for interpreting the departure of the mean profiles from a monochromatic shape.

\begin{figure}
\centering
\includegraphics[width=0.99\textwidth]{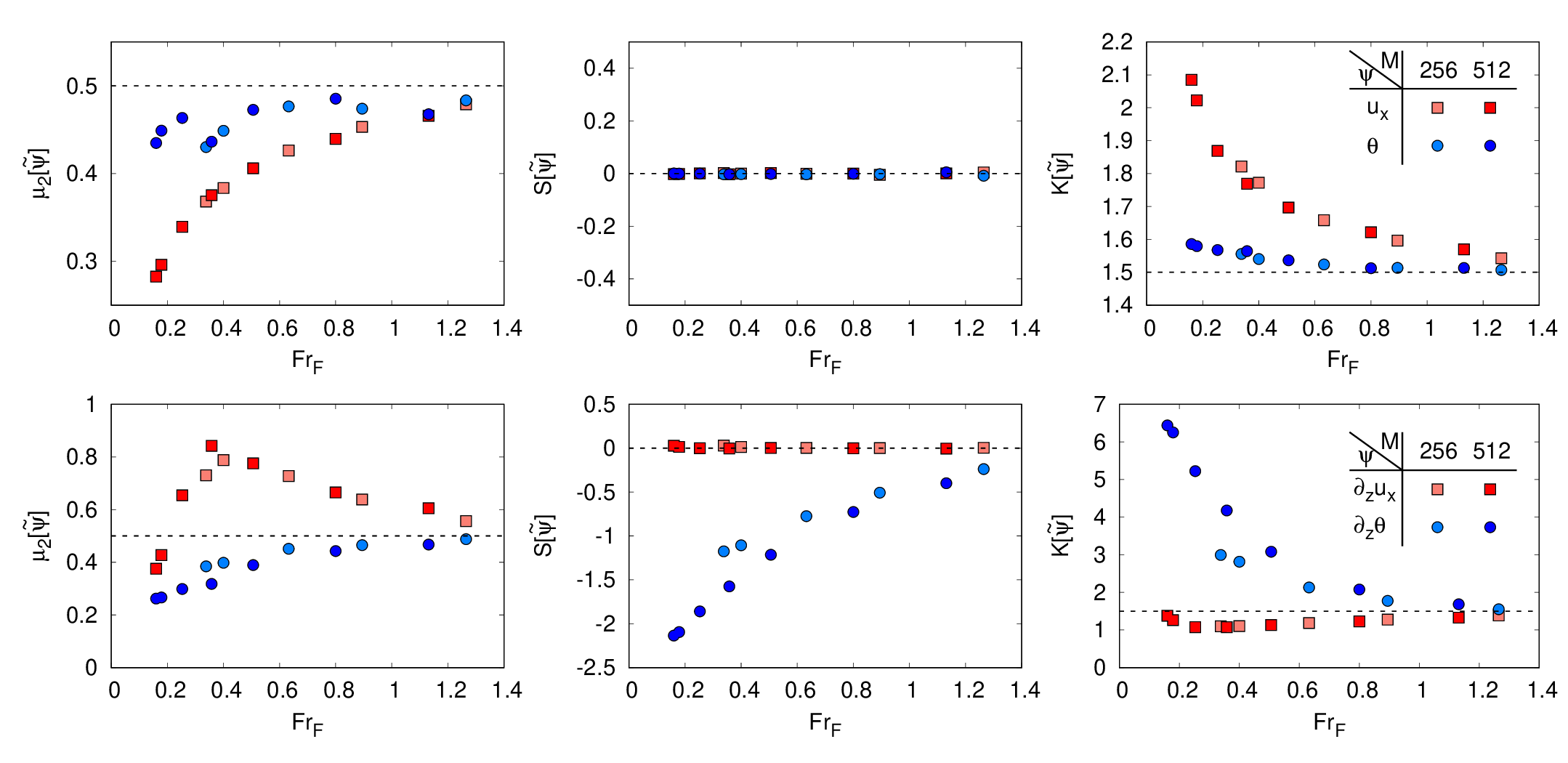}
\caption{Statistical moments of the mean profiles.  
Top row: mean velocity $\overline{u}(z)$ and mean density $\overline{\theta}(z)$.  
Bottom row: vertical gradients $d_z \overline{u}$ and $d_z \overline{\theta}$.  
For each quantity, the plots report the normalized second moment $\mu_2$ (i.e. the monochromaticity index), the skewness $\mathcal{S}$, and the kurtosis $\mathcal{K}$ of the normalized profiles, for all stratification levels, expressed in terms of $Fr_F$, and both resolutions ($M=256,512$).}
\label{fig:mono}
\end{figure}

Figure \ref{fig:mono} shows the results for the mean longitudinal velocity $\overline{u}$, the density $\overline{\theta}$ profiles (upper row), as well as their vertical derivatives 
$d_z \overline{u}$ and $d_z \overline{\theta}$ (lower row). 
For each quantity, the plots show the normalized second moment $\mu_2$ (the monochromaticity index), the skewness $\mathcal{S}$, and the kurtosis $\mathcal{K}$ of the normalized profiles, 
for all stratification levels, expressed in terms of $Fr_F$, and for both resolutions ($M=256,512$). In the unstratified case ($N=0$ and $Fr=\infty$) the profiles are sinusoidal, with $\mu_2 = 1/2$, zero skewness, and kurtosis converging to $3/2$. A reminiscence of this behaviour remains at weak stratification ($Fr \ge 1$). We also note that results from both resolutions ($M=256,512$) generally show a reasonably good collapse onto the same curve.

As stratification increases, velocity profiles change shape, becoming steeper and approaching a piecewise-linear shape with nearly linear ramps. In this idealized limit, the variance tends to $1/3$. A similar trend occurs in the density profile $\overline{\theta}(z)$, which moves away from the sinusoidal form of the weakly stratified regime and gradually develops a layered structure. Weakly stratified layers alternate with sharp interfaces, producing a staircase-like pattern (see Figures~\ref{fig:sect_te_film} and \ref{fig:profs}). As discussed in \S\ref{subsec:interface}, the asymmetry between mixed layers and interfaces plays a key role in the emergence of staircase structures in the density field.

Additional insight comes from the statistical moments of vertical gradients $d_z \overline{u}$ and $d_z \overline{\theta}$. The velocity gradient shows a non-monotonic behaviour at both weak and strong stratification. For the density gradient, the skewness is negative, consistent with the asymmetry observed in the density field in homogeneous stratified turbulence \citep[][]{kimura2016,maffioli2019,kimura2024}.

\bibliographystyle{jfm}
\bibliography{biblio}

\end{document}